%% file: K-user_MISO_BC_with_Hybrid_CSIT__v4_.tex
\documentclass[12pt,english,onecolumn,draftnocls]{IEEEtran}
\usepackage[T1]{fontenc}
\usepackage[latin9]{inputenc}
\usepackage{babel}
\usepackage{amsthm}
\usepackage{amsmath}
\usepackage{amssymb}
\usepackage{graphicx}
\usepackage{setspace}
\PassOptionsToPackage{normalem}{ulem}
\usepackage{ulem}
\usepackage[unicode=true]
 {hyperref}

\makeatletter
  \theoremstyle{plain}
  \newtheorem{thm}{\protect\theoremname}
  \theoremstyle{remark}
  \newtheorem{rem}{\protect\remarkname}
  \theoremstyle{plain}
  \newtheorem{cor}{\protect\corollaryname}

\makeatother

\providecommand{\corollaryname}{Corollary}
\providecommand{\remarkname}{Remark}
\providecommand{\theoremname}{Theorem}

\begin{document}

\title{On the DoF Region of the K-user MISO Broadcast Channel with Hybrid
CSIT}

\author{Kaniska Mohanty and Mahesh K. Varanasi~%
\thanks{The authors are with the Department of Electrical, Computer, and Energy
Engineering, University of Colorado, Boulder, CO 80309-0425, e-mail:
\protect\href{mailto:kaniska.mohanty@colorado.edu,  varanasi@colorado.edu}{kaniska.mohanty@colorado.edu,  varanasi@colorado.edu}.%
}}
\maketitle
\begin{abstract}
An outer bound for the degrees of freedom (DoF) region of the K-user
multiple-input single-output (MISO) broadcast channel (BC) is developed
under the hybrid channel state information at transmitter (CSIT) model,
in which the transmitter has instantaneous CSIT of channels to a subset
of the receivers and delayed CSIT of channels to the rest of the receivers.
For the 3-user MISO BC, when the transmitter has instantaneous CSIT
of the channel to one receiver and delayed CSIT of channels to the
other two, two new communication schemes are designed, which are able
to achieve the DoF tuple of $\left(1,\frac{1}{3},\frac{1}{3}\right)$,
with a sum DoF of $\frac{5}{3}$, that is greater than the sum DoF
achievable only with delayed CSIT. Another communication scheme showing
the benefit of the alternating CSIT model is also developed, to obtain
the DoF tuple of $\left(1,\frac{4}{9},\frac{4}{9}\right)$ for the
3-user MISO BC.\end{abstract}

\section{Introduction}

The degrees of freedom (DoF) region, defined as the pre-log factor
of the capacity in the high SNR regime, of the broadcast channel (BC)
has received considerable attention in the past few years. Under the
idealized assumption of perfect CSIT , the capacity of the Gaussian
multiple input multiple output (MIMO) BC was obtained in \cite{Weingarten2006},
which showed that the sum-DoF of a MIMO BC with $M$ antennas at the
transmitter and $N_{i}$ antennas at the $i^{\text{th}}$ receiver
is $\min\left(M,\sum_{i}N_{i}\right)$. Later, the DoF region of the
MIMO BC with no CSIT was characterized for the 2-user MIMO BC with
i.i.d. Rayleigh fading in \cite{Huang2012} and for the K-user MIMO
BC for a general class of fading distributions (that includes i.i.d.
Rayleigh fading as well as the more general isotropic fading model)
under which the transmit directions to different receive antennas
are statistically indistinguishable in \cite{Vaze2012}, where it
was established that the complete absence of any channel state information
at the transmitter leads to a collapse of the DoF to that achievable
through only time-sharing.

Investigations by Maddah-Ali and Tse in \cite{Maddah-Ali2012}, into
the DoF region of the MISO BC under the assumption of delayed CSIT,
showed that even delayed CSIT can lead to significant gains in DoF
beyond that with no CSIT, through the exploitation of the side information
available at the various receivers. While further extensions of this
result to the MIMO BC have been made, notably in \cite{Vaze2011}
and \cite{Abdoli2011}, it has also sparked interest in other CSIT
models, which combine both perfect CSIT and delayed CSIT in various
ways. The mixed CSIT model was investigated in \cite{Yang2012,Gou2012,Chen2012},
where the transmitter has the same combination of delayed CSIT and
an imperfect version of the instantaneous CSIT from each receiver,
and the DoF region for the 2-user MISO BC was completely characterized
under this model. The alternating CSIT model was introduced in \cite{Tandon2012},
where the CSIT state for each receiver is allowed to alternate between
no CSIT, delayed CSIT and instantaneous CSIT states, and the DoF region
of the 2-user MISO BC was characterized under this model in \cite{Tandon2012}. 

The DoF region of the 2-user MIMO BC and the 2-user MIMO IC were respectively
characterized in \cite{Tandon2012a} and \cite{Mohanty2012}, under
the hybrid CSIT model, where the CSIT about one receiver is delayed
and the CSIT about the other receiver is instantaneous. In this paper,
we prove an outer bound for the DoF region of the K-user MISO BC in
its most general hybrid CSIT setting, with an arbitrary number of
receivers in the instantaneous CSIT state and the rest of the receivers
in the delayed CSIT state, and then apply this outer bound to the
3-user MISO BC, where the transmitter has instantaneous CSIT about
one receiver and delayed CSIT about the other two. Although we are
unable to prove that this outer bound is tight, we provide two communication
schemes for the 3-user MISO BC which achieve a DoF tuple of $\left(1,\frac{1}{3},\frac{1}{3}\right)$
for a total sum-DoF of $\frac{5}{3}$, which exceeds the sum-DoF of
$\frac{18}{11}$ achievable with only delayed CSIT. We also illustrate
another communication scheme that uses alternating CSIT to achieve
the DoF tuple $\left(1,\frac{4}{9},\frac{4}{9}\right)$, which is
a corner point of the DoF outer bound region derived for the hybrid
CSIT model.

The DoF region of the K-user MISO BC under the alternating CSIT model
was also independently investigated by \cite{Rassouli2013}, as was
discovered by the authors after the completion of this work. 

Although the hybrid CSIT model is applied to the 2-user MIMO BC in
\cite{Tandon2012a}, the 3-user MISO BC problem is more difficult.
To tackle this problem, the communication schemes developed in this
paper for the 3-user BC illustrate two new ideas, \emph{layer peeling}\textbf{
}and \emph{serial interference alignment}. In layer peeling, we create
a structured auxiliary interference symbol which is a combination
of two other interference symbols, such that this auxiliary symbol
is useful in its entirety at one receiver, which does not need to
decipher the constituent symbols, while simultaneously allowing another
receiver to use the layered structure to peel off an already known
layer and access the constituent interference symbols. Its usefulness,
as is apparent from the communication schemes, lies in simultaneously
providing detailed information to one receiver while hiding unnecessary
details from another receiver. The communication schemes in this paper
also utilize the idea of serial interference alignment, a scenario
in which there exists a series of interference alignment by interference
symbols, with each aligned interferencesymbol allowing the receiver
to decipher the next symbol in the series, culminating in the decoding
of a desired symbol.

The channel model is given in the next section, followed by theorems
and proofs of the K-user and 3-user outer bounds, in Section \ref{sec:Outer-Bound}.
Section \ref{sec:Achievability-Schemes} explains all the communication
schemes in detail, and the paper concludes with Section \ref{sec:Conclusion}.

\section{Channel Model\label{sec:Channel-Model}}

 We start with the hybrid CSIT model for the the 3-user MISO BC,
before generalizing it to the K-user MISO BC. In the hybrid CSIT model
considered here for the 3-user BC, transmitter $T$ has $3$ transmitting
antennas, and the $3$ receivers $R_{1},R_{2}$ and $R_{3}$ have
a single antenna each. The received outputs at the receivers $R_{1},R_{2}$
and $R_{3}$ are given, respectively, by the following equations :
\begin{eqnarray*}
Y_{1}(t) & = & H_{1}(t)X(t)+Z_{1}(t),\\
Y_{2}(t) & = & H_{2}(t)X(t)+Z_{2}(t).\\
Y_{3}(t) & = & H_{3}(t)X(t)+Z_{3}(t),
\end{eqnarray*}
where, at time $t$, $Y_{i}(t)\in\mathbb{C}^{1\times1}$ is the output
at receiver $R_{i}$, $X(t)\in\mathbb{C}^{3\times1}$ is the transmitted
signal, $Z_{i}(t)$ is the additive complex Gaussian noise at $R_{i}$,
and $H_{i}(t)\in\mathbb{C}^{1\times3}$ is the channel from the transmitter
to receiver $R_{i}$, $i\in\{1,2,3\}$. The transmitted signal has
a power constraint of $P$ i.e $E(||X_{i}(t)||^{2})\leq P.$ The channel
matrices $H_{i}(t)$ are assumed to be i.i.d. over time and independent
of each other. Since, additive noise does not affect the DoF region,
we disregard the noise henceforth.

The main scope of this paper is the investigation of the DoF region
of the 3-user BC specified above, under the\textbf{ }following CSI
assumptions: 
\begin{itemize}
\item Transmitter $T$ learns the channels $H_{2}(t)$ and $H_{3}(t)$ with
a unit delay i.e., $H_{2}(t)$ and $H_{3}(t)$ are known at the transmitter
only at time $t+1$.
\item Receivers have global CSI i.e., all receivers know \emph{all} channels.
\end{itemize}
The above CSIT model is known as the \emph{hybrid CSIT}\textbf{ }model.

Let $\mathcal{M}_{1}$,$\mathcal{M}_{2}$ and $\mathcal{M}_{3}$ be
the three independent messages to be sent from the transmitter to
$R_{1}$, $R_{2}$ and $R_{3}$ respectively. A rate tuple $\left(\mathcal{R}_{1}(P),\mathcal{R}_{2}(P),\mathcal{R}_{3}(P)\right)$
is said to be achievable if there exists a codeword spanning $n$
channel uses, with a power constraint of $P$, such that the probability
of error at all receivers goes to zero as $n\rightarrow\infty$, where
$R_{i}(P)=\log(|\mathcal{M}_{i}|)/n$. The capacity region $(\mathcal{C}(P))$
of the BC is the region of all such achievable rate tuples, and the
DoF is defined as the pre-log factor of the capacity region as $P\rightarrow\infty$
i.e., 
\[
D=\biggl\{(d_{1},d_{2},d_{3})\biggl|\ d_{i}\geq0\ {\rm and}\ \exists\ \left((\mathcal{R}_{1}(P),\mathcal{R}_{2}(P),\mathcal{R}_{3}(P)\right)\in\mathcal{C}(P)
\]
\[
\left.\text{\text{such that}}\ d_{i}=\lim_{P\rightarrow\infty}\frac{\mathcal{R}_{i}(P)}{\log(P)}\:,i\in\{1,2,3\}\right\} .
\]
This notion can be extended to multiple-order messages i.e., common
messages intended for multiple receivers, and multiple-order DoF can
consequently be defined, following the approach in \cite{Maddah-Ali2012}.
Let $\mathcal{M}_{ij}$ (with $\left(i,j\right)\in\left\{ \left(1,2\right),\left(1,3\right),\left(2,3\right)\right\} $)
be an order-2 message intended for both $R_{i}$ and $R_{j}$, at
a rate of $\mathcal{R}_{ij}(P)$. We define the order-2 DoF $d_{ij}$
for the receiver pair $R_{i},R_{j}$ as 
\[
d_{ij}=\lim_{P\rightarrow\infty}\frac{\mathcal{R}_{ij}(P)}{\log(P)},\,\left(i,j\right)\in\left\{ \left(1,2\right),\left(1,3\right),\left(2,3\right)\right\} .
\]
The order-3 DoF $d_{123}$ is similarly defined, as the common DoF
for the receivers $\left(R_{1},R_{2},R_{3}\right)$.

We now define the hybrid CSIT model for the K-user MISO BC, where
the transmitter has $K$ antennas and each of the receivers $R_{i}$,
$i\in\{1,\dots,K\}$ has a single antenna, with the following CSI
assumptions:
\begin{itemize}
\item Transmitter $T$ learns the channels to the first (WLOG) $K_{P}$
receivers instantaneously i.e., $H_{1}(t),\dots,H_{K_{P}}(t)$ are
known at the transmitter at time $t$.
\item Transmitter $T$ learns the channels to the remaining $K-K_{P}$ receivers
with unit delay i.e., $H_{K_{P}+1}(t),\dots,H_{K}(t)$ are known at
the transmitter only at time $t+1$.
\item Receivers have global CSI i.e., all receivers know \emph{all }channels.
\end{itemize}
Global CSI at the receivers is assumed for the remainder of the paper,
and shall not be mentioned again for the sake of brevity.

The DoF region for the K-user MISO BC is defined analogous to the
$3$-user MISO BC. Let $\mathcal{M}_{1},\dots,\mathcal{M}_{K}$ be
$K$ independent messages to be sent to $R_{1},\dots,R_{K}$ respectively.
A rate tuple $\left(\mathcal{R}_{1}(P),\mathcal{R}_{2}(P),\dots,\mathcal{R}_{K}(P)\right)$
is said to be achievable if there exists a codeword with a power constraint
$P$ spanning $n$ channel uses, such that the probability of error
goes to zero as $n\rightarrow\infty$, with $R_{i}(P)=\log(|\mathcal{M}_{i}|)/n$.
The capacity region $\mathcal{C}(P)$ is the region of all such achievable
rate tuples, and the DoF is defined as the pre-log factor of the capacity
region as $P\rightarrow\infty$ i.e., 
\[
D=\biggl\{\left(d_{i}\right)_{i=1}^{K}\biggl|\ d_{i}\geq0\ {\rm and}\ \exists\ \left((\mathcal{R}_{1}(P),\mathcal{R}_{2}(P),\dots,\mathcal{R}_{K}(P)\right)\in\mathcal{C}(P)
\]
\[
\left.\text{\text{such that}}\ d_{i}=\lim_{P\rightarrow\infty}\frac{\mathcal{R}_{i}(P)}{\log(P)}\:,i\in\{1,\dots,K\}\right\} .
\]

We also define multiple-order messages e.g., $\mathcal{M}_{\mathcal{S}}$
intended for all the users in the subset $\mathcal{S}\subseteq\left\{ 1,2,\dots,K\right\} $,
at a rate $\mathcal{R_{S}}\left(P\right)$, and the consequent multiple-order
DoF $d_{\mathcal{S}}$, in the same manner as before i.e., 
\[
d_{\mathcal{S}}=\lim_{P\rightarrow\infty}\frac{\mathcal{R_{S}}(P)}{\log(P)},\,\mathcal{S}\subseteq\left\{ 1,2,\dots,K\right\} .
\]

\section{\label{sec:Outer-Bound}Outer Bound}

\begin{figure}
\begin{center} \def\svgwidth{.8\textwidth} 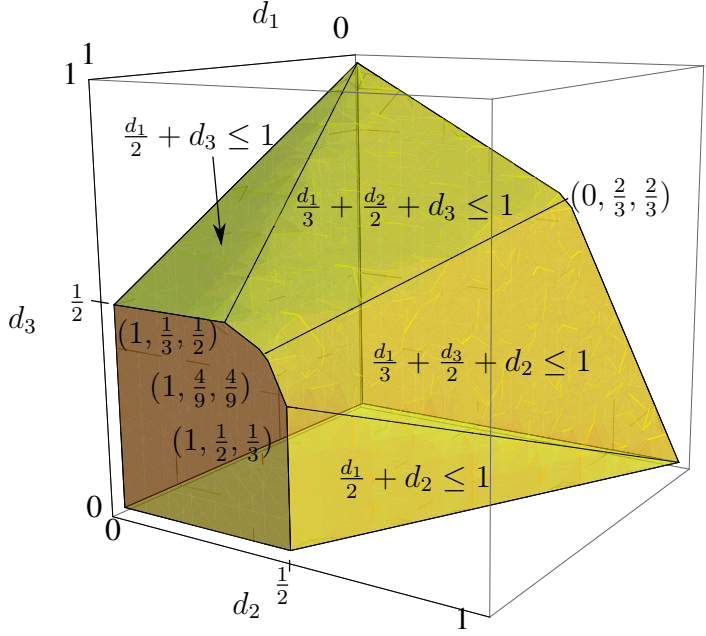 \end{center}

\caption{\label{fig:Complete-outer-bound}The $\left(d_{1},d_{2},d_{3}\right)$
outer bound for the DoF region of the 3-user MISO BC under the Hybrid
CSIT model. }
\end{figure}

\begin{figure}
\includegraphics{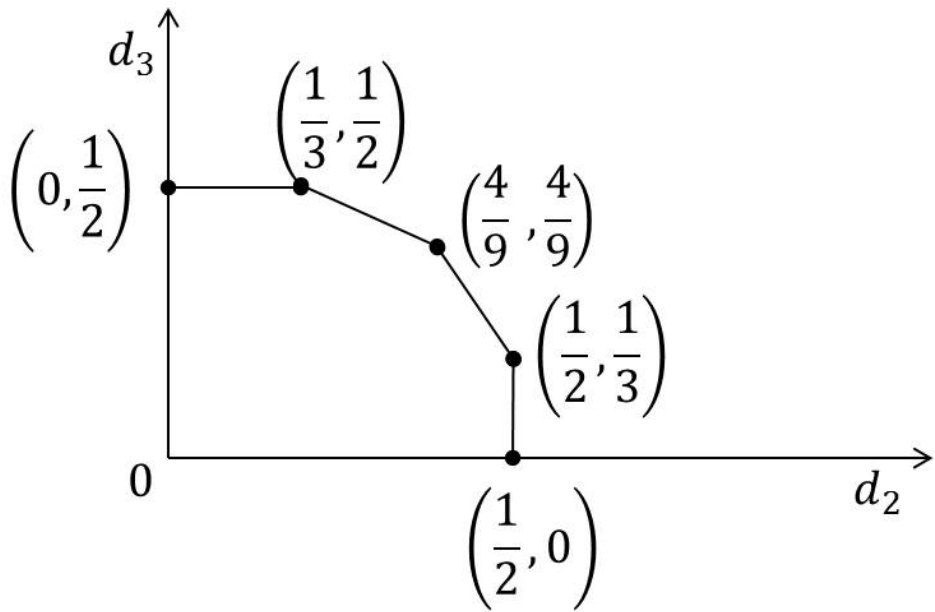}

\caption{\label{fig:The--DoF-region}The $(d_{2},d_{3})$ DoF outer bound,
in the $d_{1}=1$ plane, of the 3-user MISO BC under the Hybrid CSIT
model.}
\end{figure}

We define the notation for the ensuing theorem here. Of the $K$ receivers
in the BC, $\mathcal{E}_{P}:=\left\{ 1,2,\dots,K_{P}\right\} $ is
the set of all users about whose channels the transmitter has instantaneous
CSIT.Now, $\left\{ K_{P}+1,\dots,K\right\} $ are the remaining users
about whose channels the transmitter has only delayed CSIT, and $\mathcal{E}_{D}$
is defined to be any non-empty subset of this set i.e., $\mathcal{E}_{D}\subseteq\left\{ K_{P}+1,\dots,K\right\} ,\,\mathcal{E}_{D}\neq\phi$.
We stress a subtle notational incongruity here; while $\mathcal{E}_{P}$
is defined as the set of \emph{all} receivers for which the transmitter
has instantaneous CSIT, $\mathcal{E}_{D}$ is defined as a non-empty
subset of the set of all receivers with delayed CSIT. Also, we shall
use $\mathcal{E}_{P}$ to denote both the set of users $\left\{ 1,\dots,K_{P}\right\} $
and their corresponding receivers $\left\{ R_{1},\dots,R_{K_{P}}\right\} $,
with a similar abuse of notation for $\mathcal{E}_{D}$ and $\mathcal{S}\subseteq\left\{ 1,2,\dots,K\right\} $.
We trust that the context will make the usage clear, without causing
any confusion. $\pi_{P}$ and $\pi_{D}$ are permutation functions,
that permute the sets $\mathcal{E}_{P}$ and $\mathcal{E}_{D}$ separately.
The sequences of users in $\mathcal{E}_{P}$ and $\mathcal{E}_{D}$
after applying permutation $\pi_{P}$ and $\pi_{D}$ respectively
are $\left\{ \pi_{P}\left(1\right),\pi_{P}\left(2\right),\dots,\pi_{P}\left(K_{P}\right)\right\} $
and $\left\{ \pi_{D}\left(1\right),\pi_{D}\left(2\right),\dots,\pi_{D}\left(\left|\mathcal{E}_{D}\right|\right)\right\} $.
For any subset $\mathcal{S}$ of receivers, the multiple-order DoF
$d_{\mathcal{S}}$ has already been defined in Section \ref{sec:Channel-Model}. 
\begin{thm}
An outer-bound for the DoF region of the K-user MISO BC with hybrid
CSIT, with a combination of private (order-1) and common (multiple-order)
messages, is \label{thm:K-user-outer-bound}
\begin{multline}
\text{D}_{\text{outer}}^{\text{h-CSIT}}=\left\{ \left(d_{\mathcal{S}}\right)_{\mathcal{S}\subseteq\left\{ 1,2,\dots,K\right\} }\left|\ 0\leq d_{S}\leq1\ \forall\mathcal{S},\ \sum_{i=1}^{K_{P}}\frac{1}{K_{P}+\left|\mathcal{E}_{D}\right|-i+1}\sum_{\begin{array}{c}
\mathcal{S}\subseteq\mathcal{E}_{P}\backslash\left\{ \pi_{P}(i+1),\dots,\pi_{P}(K_{P})\right\} \\
\pi_{P}(i)\in\mathcal{S}
\end{array}}d_{\mathcal{S}}\right.\right.\\
\left.+\sum_{i=1}^{\left|\mathcal{E}_{D}\right|}\frac{1}{\left|\mathcal{E}_{D}\right|-i+1}\sum_{\begin{array}{c}
\mathcal{S}\subseteq\mathcal{E}_{P}\cup\mathcal{E}_{D}\backslash\left\{ \pi_{D}(i+1),\dots,\pi_{D}(\left|\mathcal{E}_{D}\right|)\right\} \\
\pi_{D}(i)\in\mathcal{S}
\end{array}}d_{\mathcal{S}}\leq1,\ \forall\pi_{P},\pi_{D},\mathcal{E_{D}}\right\} \label{eq:K-user-outer-bound}
\end{multline}
, where $\pi_{P}$ is any permutation of the set $\mathcal{E}_{P}:=\left\{ 1,2,\dots,K_{P}\right\} $
and $\pi_{D}$ is any permutation of the set $\mathcal{E}_{D}\subseteq\left\{ K_{P}+1,K_{P}+2,\dots,K\right\} ,\mathcal{E}_{D}\neq\phi$. 
\end{thm}
While we provide a complete and detailed proof of the theorem later
on in this section, an explanation of the above theorem and a sketch
of the proof is necessary at this point. For each of the bounds described
in Theorem \ref{thm:K-user-outer-bound}, we consider a subset of
receivers of the original BC, which contains $\mathcal{E}_{P}$ i.e.,
all receivers for which the transmitter has instantaneous CSIT, and
a non-empty subset $\mathcal{E}_{D}$ of the receivers for which the
transmitter has delayed CSIT. The subsets $\mathcal{E}_{P}$ and $\mathcal{E}_{D}$
are separately permuted using the permutation functions $\pi_{P}$
and $\pi_{D}$ respectively, and the resultant BC is then enhanced,
by giving appropriate side information at each receiver, to create
the physically degraded BC $T\rightarrow R_{\pi_{P}\left(1\right)}\rightarrow R_{\pi_{P}\left(2\right)}\rightarrow\dots\rightarrow R_{\pi_{P}\left(K_{P}\right)}\rightarrow R_{\pi_{D}\left(1\right)}\rightarrow\dots\rightarrow R_{\pi_{D}\left(\left|\mathcal{E}_{D}\right|\right)}$.
In this physically degraded BC, message $\mathcal{M_{S}}$ is a common
message meant for a subset $\mathcal{S}\subseteq\mathcal{E}_{P}\cup\mathcal{E}_{D}$.
Let $R_{\pi\left(i^{*}\right)}$ ($\pi_{P}$ or $\pi_{D}$ as the
case might be) be the receiver in the subset $\mathcal{S}$ that is
last in the degraded chain, such that any message decodable at $R_{\pi\left(i^{*}\right)}$
is decodable by all the receivers in $\mathcal{S}$, by the degradedness
of the channel. In particular, if $\mathcal{S}$ contains any receivers
from $\mathcal{E}_{D}$, we consider the largest integer $i^{*}$
such that $R_{\pi_{D}\left(i^{*}\right)}\in\mathcal{S}$, or else
if $\mathcal{S}$ contains only receivers from $\mathcal{E}_{P}$,
we consider the largest integer $i^{*}$ such that $R_{\pi_{P}\left(i^{*}\right)}\in\mathcal{S}$,
and for simplicity, we denote this receiver as $R_{\pi\left(i^{*}\right)}$.
Because of the degradedness of the channel, any message that is decodable
at $R_{\pi\left(i^{*}\right)}$ is decodable by all the receivers
in the set $\mathcal{S}$, thus allowing us to convert the common
message $\mathcal{M}_{\mathcal{S}}$ into a private message for the
receiver $R_{\pi\left(i^{*}\right)}$. For example, receiver $R_{\pi_{D}\left(\left|\mathcal{E}_{D}\right|\right)}$
decodes all messages $\mathcal{M_{S}}$ such that $R_{\pi_{D}\left(\left|\mathcal{E}_{D}\right|\right)}\in\mathcal{S}$
and $\mathcal{S}\subseteq\mathcal{E}_{P}\cup\mathcal{E}_{D}$, $R_{\pi_{D}\left(\left|\mathcal{E_{D}}\right|-1\right)}$
decodes all $\mathcal{M_{S}}$ such that $R_{\pi_{D}\left(\left|\mathcal{E}_{D}\right|-1\right)}\in\mathcal{S}$
and $\mathcal{S}\subseteq\mathcal{E}_{P}\cup\mathcal{E}_{D}\backslash\left\{ \pi_{D}\left(\left|\mathcal{E}_{D}\right|\right)\right\} $,
and so on for all the other receivers. We use the fact, from \cite{Gamal1978},
that feedback does not increase the capacity of a physically degraded
BC, to remove the delayed CSIT feedback from the receivers $\mathcal{E}_{D}$,
and then use newly created auxiliary receivers (the details are postponed
until the complete proof) to remove instantaneous CSIT about the rest
of the receivers i.e, $\mathcal{E}_{P}$. Thus, we end up with a physically
degraded channel without any feedback, with the common messages being
converted into private messages as described above. It is well known
that the capacity region of a BC without feedback depends only on
the marginal distributions of its outputs, and we can thus remove
the coupling between the receivers in the enhanced BC, so that the
receiver $R_{\pi_{P}\left(i\right)},\, i\in\left\{ 1,2,\dots,K_{P}\right\} $
now has $K_{P}+\left|\mathcal{E}_{D}\right|-i+1$ antennas and receiver
$R_{\pi_{D}\left(i\right)},\, i\in\left\{ 1,2,\dots,\left|\mathcal{E}_{D}\right|\right\} $
has $\left|\mathcal{E}_{D}\right|-i+1$ antennas. A direct application
of the results from \cite{Huang2012} and \cite{Vaze2012}, for the
DoF region of the K-user MIMO BC with no CSIT, gives the inequality
shown in (\ref{eq:K-user-outer-bound}). Permuting over all the possible
subsets $\mathcal{E}_{D}$ and permutation functions $\pi_{P}$ and
$\pi_{D}$ gives the complete outer bound region in Theorem \ref{thm:K-user-outer-bound}. 
\begin{rem}
We mentioned before the incongruity in defining $\mathcal{E}_{P}$
to be the whole set receivers with instantaneous CSIT while defining
$\mathcal{E}_{D}$ to be a subset of the receivers with delayed CSIT.
While it is possible to derive additional inequalities similar to
the ones shown in Theorem \ref{thm:K-user-outer-bound}, by considering
only a subset of the receivers with instantaneous CSIT instead of
the whole set, a careful check will show those inequalities to be
redundant. In fact, all such inequalities can be derived from the
ones we already have in Theorem \ref{thm:K-user-outer-bound}, by
substituting zero for the appropriate DoF symbols.
\end{rem}
We now illustrate the kind of DoF region obtained by applying Theorem
\ref{thm:K-user-outer-bound} to the 3-user MISO BC with hybrid CSIT.
In this model, defined previously in Section \ref{sec:Channel-Model},
the transmitter knows the channel to user $1$ instantaneously and
the channels to users $2$ and $3$ with a unit delay. We now obtain
the following corollary.
\begin{cor}
\label{thm:The-DoF-region-multi-order-symbols}The DoF region of the
3-user MISO BC with hybrid CSIT, with a combination of private (order-1)
and common (order-2 and order-3) messages, is bounded by the following
inequalities : 
\begin{eqnarray}
0\leq d_{1} & \leq & 1,\label{eq:1}\\
0\leq d_{2} & \leq & 1,\\
0\leq d_{3} & \leq & 1,\label{eq:3}\\
0\leq d_{12},d_{13},d_{23},d_{123} & \leq & 1,\\
\frac{d_{1}}{2}+d_{12}+d_{2} & \leq & 1,\label{eq:pair-begin}\\
\frac{d_{1}}{2}+d_{13}+d_{3} & \leq & 1,\label{eq:pair-end}\\
\frac{d_{1}}{3}+\frac{d_{12}+d_{2}}{2}+d_{123}+d_{13}+d_{23}+d_{3} & \leq & 1,\label{eq:converseproof}\\
\frac{d_{1}}{3}+d_{123}+d_{12}+d_{23}+d_{2}+\frac{d_{13}+d_{3}}{2} & \leq & 1.\label{eq:9}
\end{eqnarray}

\end{cor}
We also observe that the conventional DoF region of the 3-user BC
with only private messages i.e., order-1 symbols, is obtained from
Theorem \ref{thm:The-DoF-region-multi-order-symbols}, by setting
order-3 symbol i.e., $d_{123}$ and order-2 symbols $d_{12},d_{13}$
and $d_{23}$ to zero. Thus, we have the following characterization
of the outer bounds for the DoF region of the 3-user MISO BC with
purely private messages.
\begin{cor}
\label{thm:The-DoF-region-order-1-symbols}The DoF region of the 3-user
MISO BC with hybrid CSIT and only private messages is bounded by the
following inequalities :
\begin{eqnarray}
d_{1} & \leq & 1,\\
d_{2} & \leq & 1,\\
d_{3} & \leq & 1,\\
\frac{d_{1}}{2}+d_{2} & \leq & 1,\\
\frac{d_{1}}{2}+d_{3} & \leq & 1,\\
\frac{d_{1}}{3}+\frac{d_{2}}{2}+d_{3} & \leq & 1,\label{eq:mistakenproof}\\
\frac{d_{1}}{3}+d_{2}+\frac{d_{3}}{2} & \leq & 1.
\end{eqnarray}
As a special case, the DoF region, with only private messages, of
the 3-user MISO BC with hybrid CSIT and the constraint $d_{1}=1$,
is bounded by the following inequalities: 
\begin{eqnarray}
d_{2} & \leq & \frac{1}{2},\\
d_{3} & \leq & \frac{1}{2},\\
\frac{d_{2}}{2}+d_{3} & \leq & \frac{2}{3},\\
d_{2}+\frac{d_{3}}{2} & \leq & \frac{2}{3}.
\end{eqnarray}

\end{cor}
The $(d_{1},d_{2},d_{3})$ outer bound is a 3-dimensional polyhedron
in $\mathbb{R}^{3}$ space, more specifically in the cubical space
bounded by the origin $\left(0,0,0\right)$ and the planes $d_{1}=1$,
$d_{2}=1$ and $d_{3}=1$, shown in Fig. \ref{fig:Complete-outer-bound}.
To simplify matters, we focus our attention on the shape of this region
in the $d_{1}=1$ plane, which is obtained by fixing $d_{1}=1$. The
shape of this region is illustrated in Fig. \ref{fig:The--DoF-region}.

\begin{rem}
Focusing on the $d_{1}=1$ plane of the DoF outer bound region translates
to giving priority to receiver $R_{1}$ over the other two receivers,
since the transmitter has the maximum information about $R_{1}$.
In other words, we bound the DoF region while providing the maximal
DoF $d_{1}=1$ to $R_{1}$, whose channel is known instantaneously
at the transmitter.
\end{rem}

\subsection*{Proof of Corollary \ref{thm:The-DoF-region-multi-order-symbols}}

To illustrate the techniques used in the proof of the converse, we
first give the proof of Corollary \ref{thm:The-DoF-region-multi-order-symbols}
for the 3-user case, and then generalize the proof to the K-user case
to prove Theorem \ref{thm:K-user-outer-bound}. We start with the
detailed proof of the inequality (\ref{eq:converseproof}) i.e., 
\[
\frac{d_{1}}{3}+\frac{d_{12}+d_{2}}{2}+d_{123}+d_{13}+d_{23}+d_{3}\leq1,
\]
and then outline how the rest of the bounds in Theorem \ref{thm:The-DoF-region-multi-order-symbols}
can be derived using a similar technique, which follows closely that
of the converse proof in \cite{Tandon2012a}.

We now create two additional auxiliary receivers $R'$ and $R''$
with $3$ and $2$ antennas respectively. The channels to these auxiliary
receivers are $H'(t)$ and $H''(t)$, and the corresponding outputs
are $Y'(t)$ and $Y''(t)$ respectively, where $Y'(t)\in\mathbb{C}^{3\times1}$,
$H'(t)\in\mathbb{C}^{3\times3}$ , $Y''(t)\in\mathbb{C}^{2\times1}$
and $H''(t)\in\mathbb{C}^{2\times3}$. The channels $H'(t)$ and $H''(t)$
are known at all the receivers, but are not known at the transmitter.
The new channels are mutually independent of each other as well as
the rest of the channels to the original users. The outputs for the
two auxiliary receivers are given by 
\begin{eqnarray*}
Y'(t) & = & H'(t)X(t)+Z'(t)\\
Y''(t) & = & H''(t)X(t)+Z''(t).
\end{eqnarray*}
The additive noise $Z'(t)$ and $Z''(t)$ are complex normal and are
independent of all other variables, and since additive noise does
not affect the DoF region, we shall ignore it henceforth. We now impose
the additional restriction on $H'(t)$ and $H''(t)$ that 
\begin{eqnarray*}
\text{span}\left(H'(t)\right) & = & \text{span}\left[H_{1}(t),H_{2}(t),H_{3}(t)\right],\\
\text{span}\left(H''(t)\right) & = & \text{span}\left[H_{2}(t),H_{3}(t)\right].
\end{eqnarray*}

We start by giving a concise description of our original BC (OBC):
\begin{itemize}
\item Channel outputs: $Y_{1}$ at $R_{1}$, $Y_{2}$ at $R_{2}$ and $Y_{3}$
at $R_{3}$.
\item CSIT: $H_{1}$is known instantaneously, $H_{2},H_{3}$ are known with
delay.
\end{itemize}
We enhance the OBC by providing side information $\left(Y_{2},Y_{3}\right)$
to $R_{1}$ and $Y_{3}$ to $R_{2}$. This creates a physically degraded
BC, described below:
\begin{itemize}
\item Channel outputs: $\left(Y_{1},Y_{2},Y_{3}\right)$ at $R_{1}$, $\left(Y_{2},Y_{3}\right)$
at $R_{2}$ and $Y_{3}$ at $R_{3}$.
\item CSIT: $H_{1}$ is known instantaneously, $H_{2},H_{3}$ are known
with delay.
\end{itemize}
In this $T\rightarrow R_{1}\rightarrow R_{2}\rightarrow R_{3}$ physically
degraded BC, we consider the common/order-2 message $\mathcal{M}_{13}$.
Because of the degraded nature of the BC, $R_{1}$ can now decode
any message that $R_{3}$ is able to decode. Thus it is sufficient
that $\mathcal{M}_{13}$ be decodable at $R_{3}$. Similarly, it is
sufficient that the common messages $\mathcal{M}_{12}$ be decipherable
at $R_{2}$ and $\mathcal{M}_{123}$ and $\mathcal{M}_{23}$ be decipherable
at receiver $R_{3}$. Hence, we can now convert the common message
$\mathcal{M}_{12}$ into a private message for $R_{2}$, and the common
messages $\mathcal{M}_{13},\mathcal{M}_{23}$ and $\mathcal{M}_{123}$
into private messages for receiver $R_{3}$.

It is known from \cite{Gamal1978} that feedback does not improve
the capacity region of a physically degraded BC, a fact which allows
us to remove the delayed feedback links from our degraded BC, to obtain
the following BC:
\begin{itemize}
\item Channel outputs: $\left(Y_{1},Y_{2},Y_{3}\right)$ at $R_{1}$, $\left(Y_{2},Y_{3}\right)$
at $R_{2}$ and $Y_{3}$ at $R_{3}$.
\item CSIT: $H_{1}$ is known instantaneously, $H_{2},H_{3}$ are \textbf{unknown}.
\end{itemize}
We now provide our auxiliary output $Y'$ to $R_{1}$ and $Y''$ to
$R_{2}$, thus enhancing our BC to:
\begin{itemize}
\item Channel outputs: $\left(Y_{1},Y_{2},Y_{3},\mathbf{Y'}\right)$ at
$R_{1}$, $\left(Y_{2},Y_{3},\mathbf{Y''}\right)$ at $R_{2}$ and
$Y_{3}$ at $R_{3}$.
\item CSIT: $H_{1}$ is known instantaneously, $H_{2},H_{3},H',H''$ are
unknown.
\end{itemize}
The constraint we imposed earlier that $\text{span}\left(H'(t)\right)=\text{span}\left[H_{1}(t),H_{2}(t),H_{3}(t)\right]$
shows that it is possible for receiver $R_{1}$ to calculate the outputs
$\left(Y_{1},Y_{2},Y_{3}\right)$ from its knowledge of $Y'$ and
$H_{1},H_{2},H_{3},H'$. Similarly, $R_{2}$ can calculate outputs
$\left(Y_{2},Y_{3}\right)$ from its knowledge of $Y''$ and $H_{2},H_{3}$
and $H''$. Thus, our enhanced BC is now equivalent to the following
BC:
\begin{itemize}
\item Channel outputs: $Y'$ at $R_{1}$, $Y''$ at $R_{2}$ and $Y_{3}$
at $R_{3}$.
\item CSIT: $H_{1}$ is known instantaneously, $H_{2},H_{3},H',H''$ are
unknown.
\end{itemize}
We now have an equivalent BC with three outputs $\left(Y',Y'',Y_{3}\right)$,
none of whose channel gains are known at the transmitter. The transmitter
thus has no need of the knowledge of $H_{1}$, and our enhanced BC
becomes:
\begin{itemize}
\item Channel outputs: $Y'$ at $R_{1}$, $Y''$ at $R_{2}$ and $Y_{3}$
at $R_{3}$.
\item CSIT: $H',H'',H_{3}$ are unknown.
\end{itemize}
Thus, in our final version, we have a $3$-user MIMO BC with \emph{no
CSIT}, $3$ antennas at the transmitter and $3$ antennas (from our
construction of $Y'$) at $R_{1}$, $2$ antennas (our construction
of $Y''$) at $R_{2}$ and $1$ antenna at $R_{3}$. Also, as per
the discussion above, in addition to their original private messages,
$T_{x}$ has a private message $\mathcal{M}_{12}$ for $R_{2}$ and
private messages $\mathcal{M}_{13},\mathcal{M}_{23}$ and $\mathcal{M}_{123}$
for $R_{3}$. From \cite{Huang2012} and \cite{Vaze2012}, we know
that the DoF region of this MIMO BC with no CSIT is given by 
\[
\frac{d_{1}}{\min\left(3,3\right)}+\frac{d_{12}+d_{2}}{\min(3,2)}+\frac{d_{123}+d_{13}+d_{23}+d_{3}}{\min(3,1)}\leq1,
\]
thus proving inequality (\ref{eq:converseproof}) i.e., 
\[
\frac{d_{1}}{3}+\frac{d_{12}+d_{2}}{2}+d_{123}+d_{13}+d_{23}+d_{3}\leq1.
\]

Similarly, by creating the physically degraded BC $T\rightarrow R_{1}\rightarrow R_{3}\rightarrow R_{2}$,
and converting the common messages $\mathcal{M}_{123},\mathcal{M}_{12},\mathcal{M}_{23}$
into private messages for $R_{2}$ and $\mathcal{M}_{13}$ into a
private messages for $R_{3}$, we prove inequality (\ref{eq:9}),
\[
\frac{d_{1}}{3}+d_{123}+d_{12}+d_{23}+d_{2}+\frac{d_{13}+d_{3}}{2}\leq1.
\]

The proof of the other inequalities in Theorem \ref{thm:The-DoF-region-multi-order-symbols}
follows a similar reasoning. Inequalities (\ref{eq:1})-(\ref{eq:3})
are just the MIMO outer bounds for a single antenna receiver, while
inequalities \ref{eq:pair-begin}-\ref{eq:pair-end} are proven by
using the above technique on the receiver pairs $\left(R_{1},R_{2}\right)$
and $\left(R_{1},R_{3}\right)$ respectively. Since the reasoning
is so similar, we do not mention it explicitly. We note that the 2-user
outer bound obtained by considering only receivers $R_{2}$ and $R_{3}$
is redundant, and is obtained from inequalities \ref{eq:converseproof}-\ref{eq:9},
by setting $d_{1},d_{12},d_{13},d_{123}=0$.

\subsection*{\textmd{\normalsize Proof of Theorem \ref{thm:K-user-outer-bound}}}

We now prove Theorem \ref{thm:K-user-outer-bound} in all its generality.
As defined in the Section (\ref{sec:Outer-Bound}), $\mathcal{E}_{P}:=\left\{ 1,2,\dots,K_{P}\right\} $
is the set of users about whom the transmitter has instantaneous CSIT,
while $\left\{ K_{P}+1,\dots,K\right\} $ are the remaining users
about whom the transmitter has only delayed CSIT available, and $\mathcal{E}_{D}$
is now defined to be any non-empty subset of this latter set i.e.,receivers
with delayed CSIT. $\pi_{P}$ and $\pi_{D}$ are permutation functions,
that permute the sets $\mathcal{E}_{P}$ and $\mathcal{E}_{D}$ respectively.
We now consider the BC consisting of receivers $\mathcal{E}_{P}\cup\mathcal{E}_{D}$
i.e., a combination of ALL receivers for which instantaneous CSIT
exists and the subset $\mathcal{E}_{D}$ of receivers for which the
transmitter has delayed CSIT, and give below a concise description
of this BC : 
\begin{itemize}
\item Channel outputs: $Y_{i}$ at $R_{i}$, $i\in\mathcal{E}_{P}\cup\mathcal{E}_{D},$
\item CSIT: $H_{i}\;\forall i\in\mathcal{E}_{P}$ known instantaneously
, $H_{j}\;\forall j\in\mathcal{E}_{D}$ known with delay.
\end{itemize}
We now consider a permutation $\pi_{P}$ of the set $\mathcal{E}_{P}$
and a permutation $\pi_{D}$ of the set $\mathcal{E}_{D}$, and enhance
the BC by providing the output of receiver $R_{\pi_{D}(i)}$ to receivers
$R_{\pi_{D}\left(i-1\right)},R_{\pi_{D}(i-2)},\dots,R_{\pi_{D}(1)},R_{\pi_{P}\left(K_{P}\right)},\dots,R_{\pi_{P}(1)}$
, $\forall i,\,1\leq i\leq\left|\mathcal{E}_{D}\right|$ , and the
output of receiver $R_{\pi_{P}(i)}$ to receivers $R_{\pi_{P}(i-1)},R_{\pi_{P}(i-2)},\dots,R_{\pi_{P}(1)}$,
$\forall i,\,1\leq i\leq K_{P}$. This creates a physically degraded
BC, described below:
\begin{itemize}
\item Channel outputs: $\left(Y_{\pi_{P}(1)},\dots,Y_{\pi_{P}(K_{P})},Y_{\pi_{D}(1)},\dots,Y_{\pi_{D}(\left|\mathcal{E}_{D}\right|)}\right)$
at $R_{\pi_{P}(1)}$,\ldots{}, $\left(Y_{\pi_{D}(1)},\dots,Y_{\pi_{D}(\left|\mathcal{E}_{D}\right|)}\right)$
at $R_{\pi_{D}(1)}$,\ldots{}, $\left(Y_{\pi_{D}(\left|\mathcal{E}_{D}\right|)}\right)$
at $R_{\pi_{D}(\left|\mathcal{E}_{D}\right|)}$,
\item CSIT: $H_{\pi_{P}\left(i\right)}\;\forall i\in\mathcal{E}_{P}$ are
known instantaneously , $H_{\pi_{D}\left(j\right)}\;\forall j\,1\leq j\leq\left|\mathcal{E}_{D}\right|$
are known with delay.
\end{itemize}
In this $T\rightarrow R_{\pi_{P}(1)}\rightarrow R_{\pi_{P}(2)}\rightarrow\dots\rightarrow R_{\pi_{P}(K_{P})}\rightarrow R_{\pi_{D}(1)}\rightarrow\dots\rightarrow R_{\pi_{D}\left(\left|\mathcal{E}_{D}\right|\right)}$
physically degraded BC, let $\mathcal{M_{S}}$ be the multiple-order
message intended for all receivers in a set $\mathcal{S}\subseteq\mathcal{E}_{P}\cup\mathcal{E}_{D}$
. If $\mathcal{S}$ contains any receivers from $\mathcal{E}_{D}$,
we consider the largest integer $i^{*}$ such that $R_{\pi_{D}\left(i^{*}\right)}\in\mathcal{S}$,
or else if $\mathcal{S}$ contains only receivers from $\mathcal{E}_{P}$,
we consider the largest integer $i^{*}$ such that $R_{\pi_{P}\left(i^{*}\right)}\in\mathcal{S}$.
For simplicity, we denote this receiver as $R_{\pi\left(i^{*}\right)}$.
Because of the degradedness of the channel, any message that is decodable
at $R_{\pi\left(i^{*}\right)}$ is decodable by all the receivers
in the set $\mathcal{S}$, thus allowing us to convert the common
message $\mathcal{M}_{\mathcal{S}}$ into a private message for the
receiver $R_{\pi\left(i^{*}\right)}$. For example, receiver $R_{\pi_{D}\left(\left|\mathcal{E}_{D}\right|\right)}$
decodes all messages $\mathcal{M_{S}}$ such that $R_{\pi_{D}\left(\left|\mathcal{E}_{D}\right|\right)}\in\mathcal{S}$
and $\mathcal{S}\subseteq\mathcal{E}_{P}\cup\mathcal{E}_{D}$, $R_{\pi_{D}\left(\left|\mathcal{E_{D}}\right|-1\right)}$
decodes all $\mathcal{M_{S}}$ such that $R_{\pi_{D}\left(\left|\mathcal{E}_{D}\right|-1\right)}\in\mathcal{S}$
and $\mathcal{S}\subseteq\mathcal{E}_{P}\cup\mathcal{E}_{D}\backslash\left\{ \pi_{D}\left(\left|\mathcal{E}_{D}\right|\right)\right\} $
and so on for all the other receivers. 

It is known from \cite{Gamal1978} that feedback does not improve
the capacity region of a physically degraded BC. This allows us to
remove the delayed feedback links from our degraded BC, obtaining
the following ``enhanced'' BC:
\begin{itemize}
\item Channel outputs: $\left(Y_{\pi_{P}(1)},Y_{\pi_{P}(2)},\dots,Y_{\pi_{P}(K_{P})},Y_{\pi_{D}(1)},\dots,Y_{\pi_{D}(\left|\mathcal{E}_{D}\right|)}\right)$
at $R_{\pi_{P}(1)}$, \ldots{}, $\left(Y_{\pi_{D}(1)},\dots,Y_{\pi_{D}(\left|\mathcal{E}_{D}\right|)}\right)$
at $R_{\pi_{D}(1)}$,\ldots{}, $\left(Y_{\pi_{D}(\left|\mathcal{E}_{D}\right|)}\right)$
at $R_{\pi_{D}(\left|\mathcal{E}_{D}\right|)}$,
\item CSIT: $H_{\pi_{P}\left(i\right)}\;\forall i\in\mathcal{E}_{P}$ known
instantaneously , $H_{\pi_{D}\left(j\right)}\;\forall j\,1\leq j\leq\left|\mathcal{E}_{D}\right|$
\textbf{unknown}.
\end{itemize}
We now create $K_{P}+\left|\mathcal{E}_{D}\right|-1$ additional auxiliary
receivers $R_{\pi_{P}\left(1\right)}^{'},\dots,R{}_{\pi_{P}\left(K_{P}\right)}^{'},R{}_{\pi_{D}\left(1\right)}^{'},\dots,R_{\pi_{D}\left(\mathcal{\left|E_{D}\right|}-1\right)}^{'}$,
the first receiver $R'_{\pi_{P}\left(1\right)}$ with $K_{P}+\left|\mathcal{E}_{D}\right|$
antennas, and each consecutive receiver having one less antenna than
its predecessor. The outputs at the receivers are denoted as $Y_{\pi_{P}(1)}^{'},\dots,Y_{\pi_{P}(K_{P})}^{'},Y_{\pi_{D}(1)}^{'},\dots,Y_{\pi_{D}\left(\left|\mathcal{E}_{D}\right|-1\right)}^{'}$,
and the corresponding channels are similarly labeled $H_{\pi_{P}(1)}^{'},\dots,H_{\pi_{P}(K_{P})}^{'},H_{\pi_{D}(1)}^{'},\dots,H_{\pi_{D}\left(\left|\mathcal{E}_{D}\right|-1\right)}^{'}$
, the time indices being suppressed for brevity. The new channels
are mutually independent of each other as well as the rest of the
channels to the original receivers, and are known instantaneously
at all the receivers, but are unknown at the transmitter. We now impose
the following restriction on each of the newly created channels : 

\begin{eqnarray}
\text{span}\left(H_{\pi_{P}(1)}^{'}(t)\right) & = & \text{span}\left[H_{\pi_{P}(1)}(t),H_{\pi_{P}(2)}(t),\dots,H_{\pi_{P}(K_{P})}(t),H_{\pi_{D}(1)}(t),\dots,H_{\pi_{D}\left(\left|\mathcal{E}_{D}\right|\right)}(t)\right],\label{eq:constraint-begin}\\
\text{span}\left(H_{\pi_{P}(2)}^{'}(t)\right) & = & \text{span}\left[H_{\pi_{P}(2)}(t),\dots,H_{\pi_{P}(K_{P})}(t),H_{\pi_{D}(1)}(t),\dots,H_{\pi_{D}\left(\left|\mathcal{E}_{D}\right|\right)}(t)\right],\\
 & \vdots\\
\text{span}\left(H_{\pi_{D}(1)}^{'}(t)\right) & = & \text{span}\left[H_{\pi_{D}(1)}(t),\dots,H_{\pi_{D}\left(\left|\mathcal{E}_{D}\right|\right)}(t)\right],\\
 & \vdots\\
\text{span}\left(H_{\pi_{D}\left(\left|\mathcal{E}_{D}\right|-1\right)}(t)\right) & = & \text{span}\left[H_{\pi_{D}\left(\left|\mathcal{E}_{D}\right|-1\right)}(t),H_{\pi_{D}\left(\left|\mathcal{E}_{D}\right|\right)}(t)\right].\label{eq:constraint-end}
\end{eqnarray}
We provide the auxiliary output $Y_{\pi_{P}(i)}^{'}(t)$ to the receiver
$\pi_{P}(i),\ \forall i\in\mathcal{E}_{P}$, and the auxiliary output
$Y_{\pi_{D}(i)}^{'}(t)$ to the receiver $R_{\pi_{D}(i)},\ \forall i\in\mathcal{E}_{D}\backslash\left\{ \pi_{D}\left(\left|\mathcal{E}_{D}\right|\right)\right\} $.
Thus, the BC is now enhanced to the following state:
\begin{itemize}
\item Channel outputs: $\left(Y_{\pi_{P}(1)},\dots,Y_{\pi_{D}(1)},\dots,Y_{\pi_{D}\left(\left|\mathcal{E}_{D}\right|\right)},\mathbf{Y_{\pi_{P}(1)}^{'}}\right)$
at $R_{\pi_{P}(1)}$, \ldots{}, $\left(Y_{\pi_{D}(1)},\dots,Y_{\pi_{D}\left(\left|\mathcal{E}_{D}\right|\right)},\mathbf{Y_{\pi_{D}(1)}^{'}}\right)$
at $R_{\pi_{D}(1)}$,\ldots{}, $\left(Y_{\pi_{D}\left(\left|\mathcal{E}_{D}\right|-1\right)},Y_{\pi_{D}\left(\left|\mathcal{E}_{D}\right|\right)},\mathbf{Y_{\pi_{D}\left(\left|\mathcal{E}_{D}\right|-1\right)}^{'}}\right)$
at $R_{\pi_{D}(1)}$,$\left(Y_{\pi_{D}\left(\left|\mathcal{E}_{D}\right|\right)}\right)$
at $R_{\pi_{D}\left(\left|\mathcal{E}_{D}\right|\right)}$,
\item CSIT: $H_{\pi_{P}\left(i\right)}\;\forall i\in\mathcal{E}_{P}$ are
known instantaneously , $H_{\pi_{D}\left(j\right)}\;\forall j\,1\leq j\leq\left|\mathcal{E}_{D}\right|$\textbf{
}unknown\textbf{, }$H_{\pi_{P}\left(i\right)}^{'}\;\forall i\in\mathcal{E}_{P}$
are \textbf{unknown},\textbf{ $H_{\pi_{D}\left(j\right)}^{'}\;\forall j\,1\leq j\leq\left|\mathcal{E}_{D}\right|-1$
}are\textbf{ unknown.}
\end{itemize}
The constraints we imposed earlier on the span of the auxiliary channels
in equations \ref{eq:constraint-begin}-\ref{eq:constraint-end} show
that the output at the $K_{p}+\left|\mathcal{E_{D}}\right|$ antennas
of $Y_{\pi_{P}\left(1\right)}^{'}$ allow receiver $\pi_{P}(1)$ to
calculate $\left(Y_{\pi_{P}(1)},\dots,Y_{\pi_{D}(1)},\dots,Y_{\pi_{D}\left(\left|\mathcal{E}_{D}\right|\right)}\right)$
from its knowledge of $Y_{\pi_{P}\left(1\right)}^{'}$ and $H_{\pi_{P}(1)}(t),H_{\pi_{P}(2)}(t),\dots,H_{\pi_{P}(K_{P})}(t),H_{\pi_{D}(1)}(t),\dots,H_{\pi_{D}\left(\left|\mathcal{E}_{D}\right|\right)}(t)$
and $H_{\pi_{P}(1)}^{'}$. A similar reasoning follows for the rest
of the receivers in the enhanced BC, and the enhanced BC takes the
form : 
\begin{itemize}
\item Channel outputs: $Y_{\pi_{P}(1)}^{'}$ at $R_{\pi_{P}(1)}$, $Y_{\pi_{P}(2)}^{'}$
at $R_{\pi_{P}(2)}$,\ldots{}, $Y_{\pi_{D}(1)}^{'}$ at $R_{\pi_{D}(1)}$,\ldots{},
$Y_{\pi_{D}\left(\left|\mathcal{E}_{D}\right|-1\right)}^{'}$ at $R_{\pi_{D}(1)}$,$Y_{\pi_{D}\left(\left|\mathcal{E}_{D}\right|\right)}$
at $R_{\pi_{D}\left(\left|\mathcal{E}_{D}\right|\right)}$,
\item CSIT: $H_{\pi_{P}\left(i\right)}\;\forall i\in\mathcal{E}_{P}$ known
instantaneously , $H_{\pi_{D}\left(j\right)}\;\forall j\,1\leq j\leq\left|\mathcal{E}_{D}\right|$\textbf{
}unknown\textbf{, }$H_{\pi_{P}\left(i\right)}^{'}\;\forall i\in\mathcal{E}_{P}$
unknown,\textbf{ $H_{\pi_{D}\left(j\right)}^{'}\;\forall j\,1\leq j\leq\left|\mathcal{E}_{D}\right|-1$
}unknown\textbf{.}
\end{itemize}
We have thus created an equivalent BC with outputs $\left\{ Y_{\pi_{P}(1)}^{'},\dots,Y_{\pi_{P}(K_{P})}^{'},Y_{\pi_{D}(1)}^{'},\dots,Y_{\pi_{D}\left(\left|\mathcal{E}_{D}\right|-1\right)}^{'},Y_{\pi_{D}\left(\left|\mathcal{E}_{D}\right|\right)}\right\} $,
none of whose channel gains are known at the transmitter. In this
setting, the transmitter thus has no need for its instantaneous knowledge
of $H_{\pi_{P}(i)},\ i\in\mathcal{E}_{P}$, thus making out enhanced
BC equivalent to the following final BC:
\begin{itemize}
\item Channel outputs: $Y_{\pi_{P}(1)}^{'}$ at $R_{\pi_{P}(1)}$, $Y_{\pi_{P}(2)}^{'}$
at $R_{\pi_{P}(2)}$,\ldots{}, $Y_{\pi_{D}(1)}^{'}$ at $R_{\pi_{D}(1)}$,\ldots{},
$Y_{\pi_{D}\left(\left|\mathcal{E}_{D}\right|-1\right)}^{'}$ at $R_{\pi_{D}(1)}$,
$Y_{\pi_{D}\left(\left|\mathcal{E}_{D}\right|\right)}$ at $R_{\pi_{D}\left(\left|\mathcal{E}_{D}\right|\right)}$,
\item CSIT: $H_{\pi_{D}\left(j\right)}^{'}\;\forall j\,1\leq j\leq\left|\mathcal{E}_{D}\right|$\textbf{
}unknown\textbf{, }$H_{\pi_{P}\left(i\right)}^{'}\;\forall i\in\mathcal{E}_{P}$
unknown.
\end{itemize}
In this final version, we have a $K_{P}+\left|\mathcal{E}_{D}\right|$-user
MIMO BC with \emph{no CSIT}, $K$ antennas at the transmitter, with
$K_{P}+\left|\mathcal{E}_{D}\right|-i+1$ antennas at receiver $\pi_{P}\left(i\right),\ i\in\mathcal{E}_{P}$
and $\left|\mathcal{E}_{D}\right|-i+1$ antennas at receiver $\pi_{D}(i),\ i\in\mathcal{E}_{D}$,
with private messages $\mathcal{M_{S}}$ as described above. From
\cite{Huang2012} and \cite{Vaze2012}, we know that the DoF region
of this MIMO BC with no CSIT is given by 

\begin{multline}
\sum_{i=1}^{K_{P}}\frac{1}{\min\left(K,K_{P}++\left|\mathcal{E}_{D}\right|-i+1\right)}\sum_{\begin{array}{c}
\mathcal{S}\subset\mathcal{E}_{P}\backslash\left\{ \pi_{P}(i+1),\dots,\pi_{P}(K_{P})\right\} \\
\pi_{P}(i)\in\mathcal{S}
\end{array}}d_{\mathcal{S}}\\
+\sum_{i=1}^{\left|\mathcal{E}_{D}\right|}\frac{1}{\min\left(K,\left|\mathcal{E}_{D}\right|-i+1\right)}\sum_{\begin{array}{c}
\mathcal{S}\subseteq\mathcal{E}_{P}\cup\mathcal{E}_{D}\backslash\left\{ \pi_{D}(i+1),\dots,\pi_{D}(\left|\mathcal{E}_{D}\right|)\right\} \\
\pi_{D}(i)\in\mathcal{S}
\end{array}}d_{\mathcal{S}}\leq1.\label{eq:K-user-seed-eqn}
\end{multline}

Allowing for all possible non-empty subsets $\mathcal{E}_{D}$ of
$\left\{ K_{P}+1,\dots,K\right\} $ and all possible permutations
$\pi_{P}$ and $\pi_{D}$ of $\mathcal{E}_{P}$ and $\mathcal{E}_{D}$,
respectively, in inequality \ref{eq:K-user-seed-eqn} gives Theorem
\ref{thm:K-user-outer-bound}.

\section{New Communication Schemes\label{sec:Achievability-Schemes}}

\subsection{\label{sub:1st-achiev-scheme-explanation}A Scheme achieving $\left(1,\frac{1}{3},\frac{1}{3}\right)$
DoF}

\begin{figure}
\includegraphics[scale=0.55]{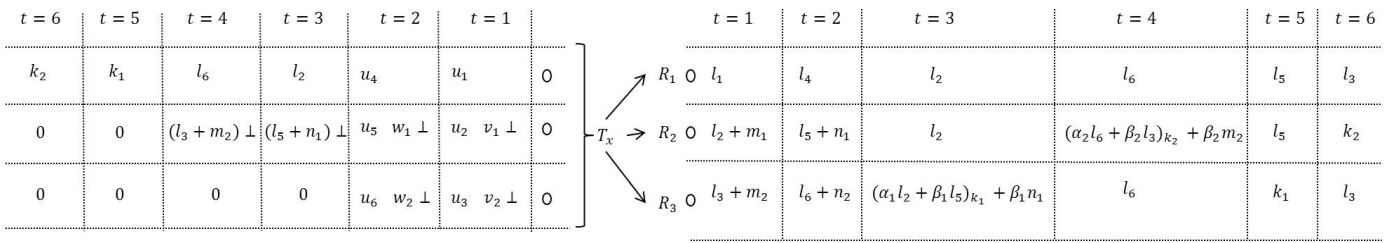}

\begin{minipage}[t]{1\columnwidth}%
\begin{center}
{\footnotesize Note :$\perp$ denotes zero-forcing at $R_{1}$ of
the adjoining data symbol.}
\par\end{center}%
\end{minipage}\caption{\label{fig:Interference-alignment-scheme-1}Interference alignment
scheme for achieving $\left(1,\frac{1}{3},\frac{1}{3}\right)$ DoF
tuple for the 3-user MISO BC under the Hybrid CSIT model. }
\end{figure}

In Fig. \ref{fig:Interference-alignment-scheme-1}, we present an
interference alignment scheme that achieves the $\left(1,\frac{1}{3},\frac{1}{3}\right)$
DoF tuple for the 3-user MISO BC under the hybrid CSIT model. We show
the achievability of this DoF tuple by coding over $6$ time slots,
during which the transmitter sends $6$ independent data symbols (DSs)
$u_{1},u_{2},u_{3},u_{4},u_{5},u_{6}$ to receiver $R_{1}$, $2$
independent DSs $v_{1}$ and $v_{2}$ to $R_{2}$ and $2$ independent
DSs $w_{1}$ and $w_{2}$ to $R_{3}$, with all the data symbols being
successfully decoded at their intended receivers. In the following
paragraphs, we explain the transmission and decoding strategy at each
time slot in detail. For the sake of clarity, we divide the complete
scheme into two phases, an initial \emph{data dissemination} phase
and a subsequent \emph{data disambiguation} phase.

\emph{\uline{Data Dissemination Phase}}

At $t=1,$ the transmitter transmits $3$ DSs $u_{1},u_{2}$ and $u_{3}$
intended for $R_{1}$ and $2$ DSs $v_{1}$ and $v_{2}$ intended
for $R_{2}$. Since the transmitter has perfect knowledge of the channel
to $R_{1}$, it utilizes this knowledge to zero-force $v_{1}$ and
$v_{2}$ at $R_{1}$. More precisely, $T_{x}$ transmits the following
signal 
\[
X(1)=\begin{bmatrix}u_{1}\\
u_{2}\\
u_{3}
\end{bmatrix}+B(1)\begin{bmatrix}0\\
v_{1}\\
v_{2}
\end{bmatrix},
\]
where $B(1)$ is the pre-coding matrix that performs transmit beamforming
in the null space of $H_{1}(1)$ i.e., $H_{1}(1)B(1)=0$. In Fig.
\ref{fig:Interference-alignment-scheme-1}, this zero-forcing at $R_{1}$
is denoted by a $\perp$ sign besides the symbol that is zero-forced,
in this case $v_{1}$ and $v_{2}$. The outputs at the three receivers
are as follows : 
\begin{eqnarray*}
Y_{1}(1) & = & H_{1}(1)\begin{bmatrix}u_{1}\\
u_{2}\\
u_{3}
\end{bmatrix}\\
 & \stackrel{\Delta}{=} & l_{1}\left(u_{1},u_{2},u_{3}\right),\\
Y_{2}(1) & = & H_{2}(1)\begin{bmatrix}u_{1}\\
u_{2}\\
u_{3}
\end{bmatrix}+H_{2}(1)B(1)\begin{bmatrix}0\\
v_{1}\\
v_{2}
\end{bmatrix}\\
 & \stackrel{\Delta}{=} & l_{2}\left(u_{1},u_{2},u_{3}\right)+m_{1}(v_{1},v_{2}),\\
Y_{3}(1) & = & H_{3}(1)\begin{bmatrix}u_{1}\\
u_{2}\\
u_{3}
\end{bmatrix}+H_{3}(1)B(1)\begin{bmatrix}0\\
v_{1}\\
v_{2}
\end{bmatrix}\\
 & \stackrel{\Delta}{=} & l_{3}\left(u_{1},u_{2},u_{3}\right)+m_{2}(v_{1},v_{2}).
\end{eqnarray*}

At $t=2$, the transmitter transmits DSs $u_{4},u_{5}$ and $u_{6}$
intended for $R_{1}$ and $w_{1},w_{2}$ intended for $R_{3}$, again
using a zero-forcing transmission strategy similar to the one used
in the previous time slot. $w_{1}$ and $w_{2}$ are zero-forced at
$R_{1}$ using a beamforming matrix $B(2)$, such that $H_{1}(2)B(2)=0$.
The outputs at the three receivers are as follows: 
\begin{eqnarray*}
Y_{1}(2) & = & l_{4}(u_{4},u_{5},u_{6}),\\
Y_{2}(2) & = & l_{5}(u_{4},u_{5},u_{6})+n_{1}(w_{1},w_{2}),\\
Y_{3}(2) & = & l_{6}(u_{4},u_{5},u_{6})+n_{2}(w_{1},w_{2}).
\end{eqnarray*}

We note here that Fig. \ref{fig:Interference-alignment-scheme-1}
depicts the same linear combinations (LCs) $l_{1},...,l_{6}$ , $m_{1},m_{2}$
and $n_{1},n_{2}$, without mentioning their underlying DSs $u_{i}$,
$v_{i}$ and $w_{i}$ respectively. By creating the convention that
the symbol $l$ e.g. $l_{1},l_{2}$ etc. shall henceforth denote linear
combinations of DSs $u_{i}$; $m_{1},m_{2}$ shall denote linear combinations
of $v_{i}$ and $n_{1},n_{2}$ shall denote linear combinations of
$w_{i}$, we dispense with the need to mention the underlying DSs
while mentioning the relevant linear combinations. We stress the fact
that all the linear combinations are almost surely independent, owing
to the generic and independent nature of the channel matrices assumed
at the outset. For the sake of brevity, we will not mention the channel
or beamforming matrices henceforth, and directly deal with the linear
combinations created at the receivers.

At the end of the Data Flooding phase,there are $6$ LCs i.e., $l_{1},...,l_{6}$
of the $6$ DSs $u_{1},...,u_{6}$ intended for $R_{1}$, of which
$R_{1}$ has knows only $l_{1}$ and $l_{4}$, without any interference.
Thus, $R_{1}$ needs to obtain the remaining four LCs in the next
phase, so that it can have six independent linear combinations of
$u_{1},...,u_{6}$, which would be sufficient to decode all of its
intended DSs. Sending the LCs $l_{2},l_{3},l_{5}$ and $l_{6}$ to
$R_{1}$ shall therefore be one of the goals of the next phase. 

The situation at $R_{2}$ is a bit more complicated. Of the two LCs
$m_{1}$ and $m_{2}$ that $R_{2}$ can use to decode its intended
DSs, $R_{2}$ observes only $m_{1}$ in this phase, at $t=1$, but
with an added interference $l_{2}$ . Thus, if $R_{2}$ learns the
LC $l_{2}$ in the next phase, it can cancel out this interference
and consequently obtain its desired LC $m_{1}$. It is to be noted
that the LC $l_{2}$ therefore plays a dual role, helping $R_{1}$
to decode its intended DSs while simultaneously aiding $R_{2}$ to
cancel out interference from its received signal. $R_{2}$ also needs
to be learn its second LC $m_{2}$ in the next phase. Thus, another
goal of the next phase shall be to ensure that $R_{2}$ learns both
$l_{2}$ and $m_{2}$. Similarly, in the next phase, $R_{3}$ needs
to learn $l_{6}$ (to cancel out the interference from $Y_{3}(2)$
and obtain $n_{2}$) as well as the LC $n_{1}$. 

\emph{\uline{Data Disambiguation Phase}}

At $t=3$, the transmitter sends two symbols $l_{2}$ and $l_{5}+n_{1}$
(the knowledge of which it has at $t=3$ due to delayed CSIT of the
channel to $R_{2}$ at times $1$ and $2$). $l_{5}+n_{1}$ is transmitted
in the null space of the channel to $R_{1}$ at $t=3$ (made possible
by the instantaneous CSIT from $R_{1}$), thus allowing $R_{1}$ to
learn $l_{2}$ without any interference. $R_{2}$ gets a linear combination
of $l_{2}$ and $l_{5}+n_{1}$, from which it cancels the contribution
of $l_{5}+n_{1}$(of which it has prior knowledge from $t=2$) and
acquires $l_{2}$. $R_{3}$, on the other hand, sees a linear combination
of $l_{2}$ and $l_{5}+n_{1},$ from which we group the interference
due to user 1 DSs into the the auxiliary symbol $k_{1}=\alpha_{1}l_{2}+\beta_{1}l_{5}$
(with $\alpha_{1}$ and $\beta_{1}$ depending on the channel from
transmitter to $R_{3}$ at $t=3$), so that $Y_{3}(3)=k_{1}+\beta_{1}n_{1}$,
as shown in Fig. \ref{fig:Interference-alignment-scheme-1}.

From our discussion earlier, we recall that $l_{2}$ is useful at
both $R_{1}$ and $R_{2}$, simultaneously providing a desired LC
to $R_{1}$ and allowing $R_{2}$ to acquire its required DS $m_{1}$
by interference cancellation. Another way of interpreting this is
to observe that $l_{2}$, while directly useful as a LC at $R_{1}$,
aligns with the interference already present at $R_{2}$ at $t=1$.
We use another layer of interference alignment to motivate our use
of the ``composite'' interference symbol $l_{5}+n_{1}$ at $t=3$.
The symbol $l_{5}+n_{1}$ provides $R_{3}$ with its desired LC $n_{1}$
(albeit with an interference $k_{1}$), and at the same time aligns
with the interference seen at $R_{2}$ at $t=2$. This interference
alignment of $l_{5}+n_{1}$ at $R_{2}$ allows $R_{2}$ to acquire
$l_{2}$ (by canceling out the interference $l_{5}+n_{1}$ at $t=3$),
which in turn is now aligned with the interference at $R_{2}$ at
$t=1$. We thus have a series of interference alignment by symbols,
in this case $l_{5}+n_{1}\rightarrow l_{2}\rightarrow m_{1}$, where
each aligned interference symbol allows the receiver (here $R_{2}$)
to decipher the next symbol in the series, which in turn is also aligned
at the same receiver, finally culminating in the decoding of a desired
symbol, here $m_{1}$, at that receiver. We call this idea \emph{serial
interference alignment}.

The idea of serial interference alignment is repeated at $t=4$, this
time focusing on aligning the interference at $R_{3}$. The transmitter
sends $l_{6}$ and $l_{3}+m_{2}$, the latter in the null space of
the channel to $R_{1}$ at $t=4.$ This allows $R_{1}$ to learn $l_{6}$
without any interference, while $R_{3}$ uses its prior knowledge
of $l_{3}+m_{2}$ (from the output at $t=1)$ to subtract out its
contribution and learn $l_{6}$. $R_{2}$ sees a linear combination
of $l_{6}$ and $l_{3}+m_{2}$, which we write in terms of the auxiliary
symbol $k_{2}=\alpha_{2}l_{6}+\beta_{2}l_{3}$ , once again combining
the interference due to user 1 DSs and $\alpha_{2}$ and $\beta_{2}$
depending on the channel to $R_{2}$ at $t=4$, so that $Y_{2}(4)=k_{2}+\beta_{2}m_{2}$.
The serial interference alignment chain $l_{3}+m_{2}\rightarrow l_{6}\rightarrow n_{2}$
allows $R_{3}$ to successfully obtain its desired LC $n_{2}$.

At time $t=5$, the transmitter sends $k_{1}$ and at time $t=2$
it sends $k_{2}$, both of which are linear combinations of symbols
of user 1. Upon receiving these, $R_{1}$ is able to cancel the contribution
of $l_{2}$ in $k_{1}$ and $l_{6}$ in $k_{2}$ to obtain the two
remaining LCs $l_{5}$ and $l_{3}$ for a total of $6$ linear combinations
$l_{1},$ \ldots{}, $l_{6}$ , thus allowing it to decode all its
DSs $u_{1},$\ldots{},$u_{6}$. Because $R_{2}$ obtains $k_{2}$
at $t=6$ it can subtract its contribution to $Y_{2}(4)$ and obtain
the second linear combination $m_{2}$ it desires, so that along with
its knowledge of $m_{1}$ it can decode $v_{1}$ and $v_{2}.$ Similarly,
by obtaining $k_{1}$ at $t=5$, $R_{3}$ can subtract its contribution
to $Y_{3}(3)$ and obtain the second linear combination $n_{1}$ it
desires so that along with its knowledge of $n_{2}$ it can decode
all its DSs $w_{1}$ and $w_{2}$.

The motivation behind grouping the interference from user 1 DSs into
auxiliary interference symbol $k_{1}$ and $k_{2}$ becomes clear
now. We observe that at $t=5$, $R_{3}$ needed only a particular
linear combination of $l_{2}$ and $l_{5}$ i.e., $k_{1}$, without
incurring the additional expense of learning $l_{2}$ and $l_{5}$
individually, to cancel out the interference $k_{1}$ from $Y_{3}(3)$
and obtain its desired LC $n_{1}$. At the same time, $R_{1}$ also
utilized $k_{1}$ by using its previous knowledge of $l_{2}$ (from
$Y_{1}(3)$) to \emph{peel off} $l_{2}$ from $k_{1}$ and access
its desired LC $l_{5}$ . Thus, the motivation behind creating the
auxiliary interference symbol i.e., $k_{1}$ was to have a structured
symbol which is useful in its entirety at one receiver ($R_{3}$),
but whose additional layered structure simultaneously allows another
receiver ($R_{1}$) to peel off already known layers ($l_{2}$) and
access more detailed useful information ($l_{5}$). We name this idea
as \emph{layer peeling}. Its usefulness, as is apparent from this
communication scheme, lies in simultaneously providing detailed information
to one receiver while hiding unnecessary details from another receiver.
Layer peeling is again used in the context of $k_{2}$ and the receivers
$R_{1}$ and $R_{2}$ at $t=6$, where $R_{1}$ peels off $l_{6}$
from $k_{2}$ to access its desired LC $l_{3}$ and $R_{2}$ uses
$k_{2}$ in its entirety to cancel out interference from $Y_{2}(4)$.
We note here that although $R_{2}$ can compute $l_{5}$ at $t=5$
and $R_{3}$ can computer $l_{3}$ at $t=6$ using layer peeling,
this knowledge has no particular use at either of the two receivers.

\subsection{Alternate Scheme for achieving $\left(1,\frac{1}{3},\frac{1}{3}\right)$
DoF}

\begin{figure}
\includegraphics[scale=0.55]{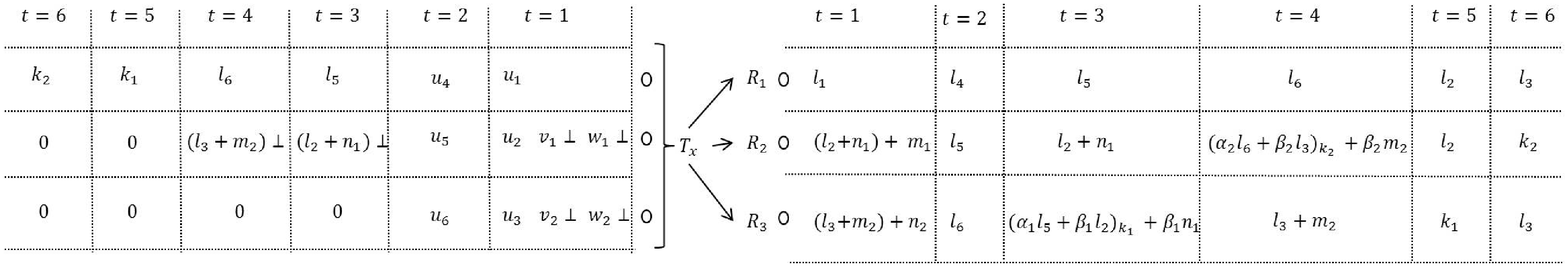}

\begin{minipage}[t]{1\columnwidth}%
\begin{center}
{\footnotesize Note :$\perp$ denotes zero-forcing at $R_{1}$ of
the adjoining data symbol.}
\par\end{center}%
\end{minipage}

\caption{\label{fig:Alternate-interference-alignment}Alternate interference
alignment scheme for achieving $\left(1,\frac{1}{3},\frac{1}{3}\right)$
DoF tuple for the 3-user MISO BC under the Hybrid CSIT model.}
\end{figure}

Fig. \ref{fig:Alternate-interference-alignment} depicts another communication
scheme for the same $\left(1,\frac{1}{3},\frac{1}{3}\right)$ DoF
tuple. Because of its similarity to the previous communication scheme,
we give here a brief description, emphasizing only the aspects in
which it differs from the former scheme. 

As before, we show the achievability of this DoF tuple by coding over
$6$ time slots, by sending $6$ independent DSs $u_{1},u_{2},u_{3},u_{4},u_{5},u_{6}$
to receiver $R_{1}$, $2$ independent DSs $v_{1}$ and $v_{2}$ to
$R_{2}$ and $2$ independent DSs $w_{1}$ and $w_{2}$ to $R_{3}$,
such that all DSs are successfully decoded at their intended receivers.
Once again, we divide the complete scheme into two phases, an initial
\emph{data dissemination} phase and a subsequent \emph{data disambiguation}
phase.

\emph{\uline{Data Dissemination Phase}}

The major difference between the present scheme and the previous one
lies in the data dissemination phase. While previously the DSs intended
for $R_{2}$ ($v_{1},v_{2}$) and the DSs intended for $R_{3}$ ($w_{1},w_{2}$)
were transmitted during separate time slots $t=1$ and $t=2$ respectively,
this scheme transmits $v_{1},v_{2},w_{1},w_{2}$ along with $u_{1},u_{2},u_{3}$
at $t=1$, thus causing a different interference pattern at the receivers.
The transmitter uses its instantaneous knowledge of the channel to
$R_{1}$ to send $v_{1},v_{2},w_{1},w_{2}$ in the null space of $H_{1}$.
More precisely, the following signal is transmitted, 
\[
X(1)=\begin{bmatrix}u_{1}\\
u_{2}\\
u_{3}
\end{bmatrix}+B(1)\begin{bmatrix}0\\
v_{1}\\
v_{2}
\end{bmatrix}+C(1)\begin{bmatrix}0\\
w_{1}\\
w_{2}
\end{bmatrix},
\]
where $B(1)$ and $C(1)$ are the pre-coding matrices that perform
transmit beamforming in the null space of $H_{1}(1)$ i.e., $H_{1}(1)B(1)=0$
and $H_{1}(1)C(1)=0$. The outputs at the three receivers are denoted
as follows:
\begin{eqnarray*}
Y_{1}(1) & = & H_{1}(1)\begin{bmatrix}u_{1}\\
u_{2}\\
u_{3}
\end{bmatrix}\\
 & \stackrel{\Delta}{=} & l_{1}\left(u_{1},u_{2},u_{3}\right),\\
Y_{2}(1) & = & H_{2}(1)\begin{bmatrix}u_{1}\\
u_{2}\\
u_{3}
\end{bmatrix}+H_{2}(1)C(1)\begin{bmatrix}0\\
w_{1}\\
w_{2}
\end{bmatrix}+H_{2}(1)B(1)\begin{bmatrix}0\\
v_{1}\\
v_{2}
\end{bmatrix}\\
 & \stackrel{\Delta}{=} & \left(l_{2}\left(u_{1},u_{2},u_{3}\right)+n_{1}(w_{1},w_{2})\right)+m_{1}(v_{1},v_{2}),\\
Y_{3}(1) & = & H_{3}(1)\begin{bmatrix}u_{1}\\
u_{2}\\
u_{3}
\end{bmatrix}+H_{3}(1)B(1)\begin{bmatrix}0\\
v_{1}\\
v_{2}
\end{bmatrix}+H_{3}(1)C(1)\begin{bmatrix}0\\
w_{1}\\
w_{2}
\end{bmatrix}\\
 & \stackrel{\Delta}{=} & \left(l_{3}\left(u_{1},u_{2},u_{3}\right)+m_{2}(v_{1},v_{2})\right)+n_{2}(w_{1},w_{2}).
\end{eqnarray*}
For further simplification, the interference terms are grouped in
brackets at the receivers $R_{2}$ and $R_{3}$ in Fig. \ref{fig:Alternate-interference-alignment}.
We stick to the previous notation of denoting the linear combinations
(LCs) of $u_{i}$ by the symbol $l$, $v_{i}$ by the symbol $m$
and $w_{i}$ by the symbol $n$ respectively. We also note that all
the linear combinations are almost surely independent, owing to the
generic and independent nature of the channel matrices.

At $t=2$, the transmitter transmits the symbols $u_{4},u_{5}$ and
$u_{6}$, thus creating the independent linear combinations $l_{4},l_{5}$
and $l_{6}$ at receivers $R_{1},R_{2}$ and $R_{3}$ respectively.
In the next phase, $R_{1}$ needs the LCs $l_{2},l_{3},l_{5}$ and
$l_{6}$, while $R_{2}$ needs to learn the LC $m_{2}$ and the composite
interference term $l_{2}+n_{1}$. Similarly, $R_{3}$ needs $n_{1}$
and the composite interference term $l_{3}+m_{2}$ in the next phase.

\emph{\uline{Data Disambiguation Phase}}

At $t=3,$ the transmitter sends two symbols $l_{5}$ and $l_{2}+n_{1}$
(the knowledge of which it has at $t=3$ due to delayed CSIT of channel
to $R_{2}$ at time 1), the latter in the null space of the channel
from the transmitter to $R_{1}$ at $t=3$ (since the receiver has
instantaneous knowledge of the channel to $R_{1}$). $R_{1}$ acquires
its desired linear combination $l_{5}$, while $R_{2}$ sees a linear
combination of $l_{5}$ and $l_{2}+n_{1}$, from which it removes
the contribution of $l_{5}$ (which it learnt at $t=2)$ to acquire
$l_{2}+n_{1}$. $R_{3}$ on the other hand observes a linear combination
of $l_{5}$ and $l_{2}+n_{1}$, which we write in terms of an auxiliary
symbol $k_{1}=\alpha_{1}l_{5}+\beta_{1}l_{2}$ (i.e., the interference
due to user 1 DSs, $\alpha_{1}$ and $\beta_{1}$ dependent on the
channel to $R_{3}$ at $t=3$) so that $Y_{3}(3)=k_{1}+\beta_{1}n_{1}$
.

The transmission strategy is similar at $t=4,$ where the symbols
$l_{6}$ and $l_{3}+m_{2}$ are transmitted, the latter in the null
space of the channel to $R_{1}$. $R_{1}$ acquires $l_{6}$, and
$R_{3}$ uses its previous knowledge of $l_{6}$ to obtain $l_{3}+m_{2}$.
The situation at $R_{2}$ parallels that of $R_{3}$ at $t=3$, where
we let $Y_{2}(4)=k_{2}+\beta_{2}m_{2}$ , with $k_{2}=\alpha_{2}l_{6}+\beta_{2}m_{2}$
(i.e., interference due to user 1's DSs with $\alpha_{2}$ and $\beta_{2}$
depending on channel from transmitter to $R_{2}$ at $t=4$).

Thus, at $t=3,4$, the transmitter sends $2$ useful linear combinations
$l_{5}$ and $l_{6}$ to $R_{1}$, the composite interference symbols
$l_{2}+n_{1}$ and $l_{3}+m_{2}$ to receivers $R_{2}$ and $R_{3}$
respectively, which then use this knowledge to cancel out the interference
from $Y_{2}(1)$ and $Y_{3}(1)$ to obtain their respective desired
linear combinations $m_{1}$ and $n_{2}$. At the same time, $R_{2}$
and $R_{3}$ are also provided with their other desired linear combination
i.e., $m_{2}$ and $n_{1}$, albeit with additional interference $k_{2}$
and $k_{1}$ respectively.

At time $t=5$ the transmitter sends $k_{1}$ and at time $t=6$ it
sends $k_{2}$. Upon receiving these, $R_{1}$ is able to cancel the
contribution of $l_{5}$ in $k_{1}$ and $l_{6}$ in $k_{2}$ to obtain
the two remaining linear combinations $l_{2}$ and $l_{3}$ for a
total of 6 linear combinations $l_{1},...,l_{6}$ so that it can obtain
its DSs $u_{1},...,u_{6}$. Because $R_{2}$ obtains $k_{2}$ at $t=6$
it can subtract its contribution to $Y_{2}(4)$ and obtain the second
linear combination $m_{2}$ it desires, so that along with its knowledge
of $m_{1}$ it can decode $v_{1}$ and $v_{2}$. Similarly, because
$R_{3}$ obtains $k_{1}$ at $t=5$ it can subtract its contribution
to $Y_{3}(3)$ and obtain the second linear combination $n_{1}$ it
desires so that along with its knowledge of $n_{2}$ it can decode
its DSs $w_{1}$ and $w_{2}$.

\subsubsection{Compactness of Communication Scheme in Fig. \ref{fig:Alternate-interference-alignment} }

We refer back to Fig. \ref{fig:The--DoF-region}, which shows $\left(1,\frac{1}{2},\frac{1}{3}\right)$
as one of the extreme points of the DoF region, which translates to
$\left(6,3,2\right)$ DSs sent over $6$ time slots. Since the communication
scheme presented in Fig. \ref{fig:Alternate-interference-alignment}
successfully sends $\left(6,2,2\right)$ DSs over $6$ time slots,
it is tempting to try and transmit an extra DS to either $R_{2}$
or $R_{3}$ using the same scheme in the hope of achieving the aforementioned
DoF extreme point. We argue in the following paragraphs against such
a possibility for the present scheme, with the help of the outer bounds
proved in Theorem \ref{thm:The-DoF-region-multi-order-symbols}.

At time $t=1,$ the transmitter transmits $3$ independent DSs intended
for $R_{1}$, which is the maximum number of independent DSs that
can be transmitted simultaneously over $3$ antennas. Visualizing
the $3$- antenna system at $T_{x}$ as a $3$- dimensional vector
space, the null space of $H_{1}$ is a $2$-dimensional subspace of
this vector space. To zero-force $v_{1}$ and $v_{2}$ (intended for
$R_{2}$) at $R_{1}$, we need to transmit the DSs using transmit
vectors in this $2$-dimensional null space. More precisely, the last
two rows of the beamforming matrix $B$ should lie in this $2$-dimensional
null space, which means that we can send a maximum of two independent
DSs in the same time slot to $R_{2}$. The same holds true for the
DSs $w_{1},w_{2}$ intended for $R_{3}$. Hence, we see that the transmission
strategy at $t=1$, under the constraint of zero-forcing $v_{i}$
and $w_{i}$ at $R_{1}$, is tight and can not be improved upon.

At the end of $t=1$, we have the following requirements for the next
$5$ time slots, $t=2,...,6$:
\begin{eqnarray*}
R_{1} & \leftarrow & l_{2},l_{3},u_{4},u_{5},u_{6},\\
R_{2} & \leftarrow & \left(l_{2}+n_{1}\right),\left(l_{3}+m_{2}\right),l_{3},\\
R_{3} & \leftarrow & \left(l_{3}+m_{2}\right),\left(l_{2}+n_{1}\right),l_{2}.
\end{eqnarray*}

We reformulate these requirements in terms of order-1 and order-2
DSs, to be sent in $5$ time slots , as 
\begin{eqnarray*}
R_{1} & \leftarrow & u_{4},u_{5},u_{6},\\
R_{1,2} & \leftarrow & l_{3},\\
R_{1,3} & \leftarrow & l_{2},\\
R_{2,3} & \leftarrow & \left(l_{2}+n_{1}\right),\left(l_{3}+m_{2}\right).
\end{eqnarray*}

We denote the cardinality of the order-1 symbols as $d_{1}^{'},d_{2}^{'},d_{3}^{'}$
and the order-2 symbols as $d_{12}^{'},d_{13}^{'},d_{23}^{'}$. From
the requirements mentioned above, we obtain the following values for
the cardinality of the various symbols:
\begin{eqnarray}
d_{1}^{'} & = & 3,\label{eq:23}\\
d_{12}^{'} & = & 1,\\
d_{13}^{'} & = & 1,\\
d_{23}^{'} & = & 2.\label{eq:26}
\end{eqnarray}

We now reformulate inequalities (\ref{eq:converseproof}) and (\ref{eq:9})
in terms of the cardinality of the symbols, keeping in mind that all
the symbols need to be transmitted in the next $5$ time slots, we
have
\begin{eqnarray*}
\frac{d_{1}^{'}}{3}+\frac{d_{12}^{'}+d_{2}^{'}}{2}+d_{13}^{'}+d_{23}^{'}+d_{3}^{'} & \leq & 5,\\
\frac{d_{1}^{'}}{3}+d_{12}^{'}+d_{23}^{'}+d_{2}^{'}+\frac{d_{13}^{'}+d_{3}^{'}}{2} & \leq & 5.
\end{eqnarray*}

Substituting the values of the cardinality from equations (\ref{eq:23})-(\ref{eq:26})
in the above inequalities, we obtain the following inequalities: 
\begin{eqnarray}
\frac{d_{2}^{'}}{2}+d_{3}^{'} & \leq & \frac{1}{2},\label{eq:27}\\
d_{2}^{'}+\frac{d_{3}^{'}}{2} & \leq & \frac{1}{2}.\label{eq:28}
\end{eqnarray}

Any attempt at sending an extra DS to $R_{2}$ in these $5$ time
slots i.e., setting $d_{2}^{'}=1$, would violate inequality (\ref{eq:28}).
Similarly, sending an extra DS to $R_{3}$ would violate inequality
(\ref{eq:27}). This proves the impossibility of sending an extra
DS to either $R_{2}$ or $R_{3}$ to achieve the DoF extreme point
$\left(1,\frac{1}{2},\frac{1}{3}\right)$ using the communication
scheme depicted in Fig. \ref{fig:Alternate-interference-alignment}.

\subsection{Communication Scheme for achieving $\left(1,\frac{4}{9},\frac{4}{9}\right)$
DoF under the Alternating CSIT Model}

\begin{figure}
\includegraphics[scale=0.55]{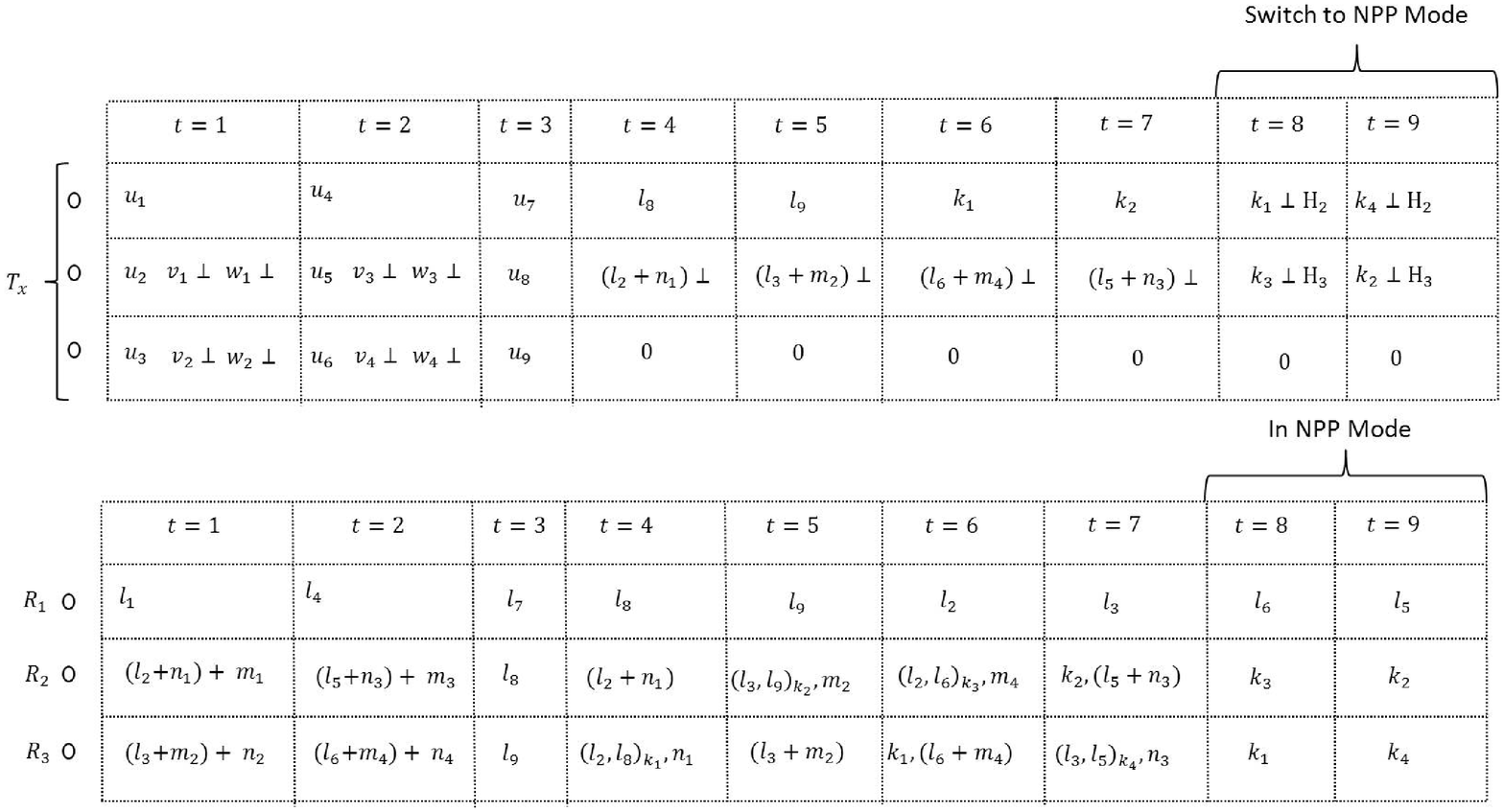}

\begin{minipage}[t]{1\columnwidth}%
\begin{center}
{\footnotesize Note :$\perp$ denotes zero-forcing at $R_{1}$ of
the adjoining data symbol.}
\par\end{center}%
\end{minipage}

\caption{ \label{fig:-Alternating-CSIT-Example}Interference alignment scheme
for achieving $\left(1,\frac{4}{9},\frac{4}{9}\right)$ DoF tuple
for the 3-user MISO BC under the Alternating CSIT model.}
\end{figure}

In Fig. \ref{fig:-Alternating-CSIT-Example}, we illustrate an communication
scheme that achieves $\left(1,\frac{4}{9},\frac{4}{9}\right)$ DoF
for the 3-user MISO BC under the alternating CSIT model, where we
show the achievability of the DoF tuple by coding over 9 time slots.
Under the alternating CSIT model considered for this particular achievable
scheme, the BC remains in the original hybrid CSIT state for $7$
time slots, during which the transmitter has instantaneous CSIT about
the channel to receiver $R_{1}$ and delayed CSIT about the channel
to receivers $R_{2}$ and $R_{3}$., and for the remaining $2$ time
slots, the transmitter switches to a new CSIT state, where it has
perfect CSIT about the channels to receivers $R_{2}$ and $R_{3}$
and no CSIT about the channel to receiver $R_{1}$, a CSIT state we
abbreviate as NPP (no CSIT, perfect CSIT, perfect CSIT). 

Over the $9$ time slots, the transmitter sends $9$ independent DSs
$u_{1},u_{2},$\ldots{}$,u_{9}$ to receiver $R_{1}$, $4$ independent
DSs $v_{1},v_{2},v_{3}$ and $v_{4}$ to $R_{2}$ and $4$ independent
DSs $w_{1},w_{2},w_{3}$ and $w_{4}$ to $R_{3}$, with all the DSs
being successfully decoded at their intended receivers, using a transmission
and decoding strategy that we explain in detail in the following paragraphs.
We divide the scheme into $3$ different phases, \emph{data dissemination
phase}\textbf{, }\emph{data disambiguation phase} and a new \emph{NPP
phase}\textbf{. }We note that in both the data dissemination and data
disambiguation phase, the transmitter is in the original hybrid state,
but switches to the NPP CSIT state in the NPP phase.

\emph{\uline{Data Dissemination Phase}}

At time $t=1$, the transmitter sends 3 DSs $u_{1},u_{2}$ and $u_{3}$
for $R_{1}$, $v_{1}$ and $v_{2}$ for $R_{2}$ and $w_{1}$ and
$w_{2}$ for $R_{3}$, using its instantaneous knowledge of the channel
to $R_{1}$ at $t=1$ to send $v_{1},v_{2},w_{1},w_{2}$ in the null
space of $H_{1}$. More precisely, the following signal is transmitted,
\[
X(1)=\begin{bmatrix}u_{1}\\
u_{2}\\
u_{3}
\end{bmatrix}+B(1)\begin{bmatrix}0\\
v_{1}\\
v_{2}
\end{bmatrix}+C(1)\begin{bmatrix}0\\
w_{1}\\
w_{2}
\end{bmatrix},
\]
where $B(1)$ and $C(1)$ are the pre-coding matrices that perform
transmit beamforming in the null space of $H_{1}(1)$ i.e., $H_{1}(1)B(1)=0$
and $H_{1}(1)C(1)=0$. In Fig. \ref{fig:Alternate-interference-alignment},
this zero-forcing is denoted by a $\perp$ sign besides the symbols
that are zero-forced at $R_{1}$. The outputs at the three receivers
are precisely as follows:
\begin{eqnarray*}
Y_{1}(1) & = & H_{1}(1)\begin{bmatrix}u_{1}\\
u_{2}\\
u_{3}
\end{bmatrix}\\
 & \stackrel{\Delta}{=} & l_{1}\left(u_{1},u_{2},u_{3}\right),\\
Y_{2}(1) & = & H_{2}(1)\begin{bmatrix}u_{1}\\
u_{2}\\
u_{3}
\end{bmatrix}+H_{2}(1)C(1)\begin{bmatrix}0\\
w_{1}\\
w_{2}
\end{bmatrix}+H_{2}(1)B(1)\begin{bmatrix}0\\
v_{1}\\
v_{2}
\end{bmatrix}\\
 & \stackrel{\Delta}{=} & \left(l_{2}\left(u_{1},u_{2},u_{3}\right)+n_{1}(w_{1},w_{2})\right)+m_{1}(v_{1},v_{2}),\\
Y_{3}(1) & = & H_{3}(1)\begin{bmatrix}u_{1}\\
u_{2}\\
u_{3}
\end{bmatrix}+H_{3}(1)B(1)\begin{bmatrix}0\\
v_{1}\\
v_{2}
\end{bmatrix}+H_{3}(1)C(1)\begin{bmatrix}0\\
w_{1}\\
w_{2}
\end{bmatrix}\\
 & \stackrel{\Delta}{=} & \left(l_{3}\left(u_{1},u_{2},u_{3}\right)+m_{2}(v_{1},v_{2})\right)+n_{2}(w_{1},w_{2}).
\end{eqnarray*}

We again use the symbol $l$ for linear combinations of $u_{i}$,
$m$ for linear combinations of $v_{i}$ and $n$ for linear combinations
of $w_{i}$. Also, all the linear combinations are almost surely linearly
independent, owing to the generic nature of the channel matrices. 

At $t=2$, the transmitter uses the exact same strategy as in the
first time slot, but now the DSs it sends are $u_{4},u_{5}$ and $u_{6}$
for $R_{1}$ and $v_{3},v_{4}$ for $R_{2}$ and $w_{3},w_{4}$ for
$R_{3}$, with $v_{3},v_{4},w_{3},w_{4}$ being transmitted in the
null space of the channel to $R_{1}$ at $t=2$. Thus, the outputs
at each of the receivers are 
\begin{eqnarray*}
Y_{1}(2) & = & l_{4}\\
Y_{2}(2) & = & \left(l_{5}+n_{3}\right)+m_{3},\\
Y_{3}(2) & = & \left(l_{6}+m_{4}\right)+n_{4},
\end{eqnarray*}
where we have grouped the interference terms separately. At $t=3$,
the transmitter transmits the remaining DSs $u_{7},u_{8}$ and $u_{9}$,
thus creating the LCs $l_{7},l_{8}$ and $l_{9}$ at $R_{1},R_{2}$
and $R_{3}$ respectively. At the end of this time slot, all the DSs
have been transmitted, and the rest of the scheme focuses on canceling
out the interference to allow each of the receivers to decode their
respective DSs. 

At the end of the data dissemination phase, of the $9$ linear combinations
$l_{1},$\ldots{}$,l_{9}$ of the DSs $u_{1}$,\ldots{},$u_{9}$,
$R_{1}$ knows only $3$ LCs and needs to acquire the other $6$ LCs
to be able to decode all its DSs . On the other hand, the $2$ LCs
$m_{1}$ and $m_{3}$ that $R_{2}$ sees comes added with the composite
interference symbols $l_{2}+n_{1}$ and $l_{5}+n_{3}$ respectively,
both of which need to be learnt in the next phase. $R_{2}$ also needs
to learn the remaining of its desired LCS i.e., $m_{2}$ and $m_{4}$.
The situation at $R_{3}$ is similar, and it needs to learn the composite
interference symbols $l_{3}+m_{2}$ and $l_{6}+m_{4}$ as well as
its own desired LCs $n_{1}$ and $n_{3}$.

\emph{\uline{Data Disambiguation Phase}}

At $t=4,$ the transmitter sends two symbols $l_{8}$ and $l_{2}+n_{1}$
(the knowledge of which it has at $t=3$ due to delayed CSIT of channel
to $R_{2}$ at time 1), the latter in the null space of the channel
from the transmitter to $R_{1}$ at $t=3$ (since the receiver has
instantaneous knowledge of the channel to $R_{1}$). $R_{1}$ acquires
its desired linear combination $l_{8}$, while $R_{2}$ sees a linear
combination of $l_{8}$ and $l_{2}+n_{1}$, from which it removes
the contribution of $l_{8}$ (which it learnt at $t=3)$ to acquire
$l_{2}+n_{1}$. We thus utilize the idea of serial interference alignment
introduced earlier, with the alignment chain being $l_{8}\rightarrow l_{2}+n_{1}\rightarrow m_{1}$,
culminating in $R_{2}$ obtaining its desired LC $m_{1}$ at the end
of the chain. $R_{3}$ on the other hand observes a linear combination
of $l_{8}$ and $l_{2}+n_{1}$, which we write in terms of an auxiliary
symbol $k_{1}=\left(l_{2},l_{8}\right)$ , where the bracket signifies
a linear combination, with the coefficients depending on the channel
to $R_{3}$ at $t=4$, so that $Y_{3}(3)=\left(k_{1},n_{1}\right)$
. To simplify notation, we shall henceforth use brackets to denote
a linear combination, the coefficients depending on the channel to
the receiver in that particular time slot at which the linear combination
is created.

The transmission strategy is similar at $t=4,$ where the symbols
$l_{9}$ and $l_{3}+m_{2}$ are transmitted, the latter in the null
space of the channel to $R_{1}$. $R_{1}$ acquires $l_{6}$, and
$R_{3}$ uses its previous knowledge of $l_{6}$ to obtain $l_{3}+m_{2}$.
The serial alignment chain at $R_{3}$ is now $l_{9}\rightarrow l_{3}+m_{2}\rightarrow n_{2}$.
The situation at $R_{2}$ parallels that of $R_{3}$ at $t=3$, where
we let $Y_{2}(4)=\left(k_{2},m_{2}\right)$ , with $k_{2}=\left(l_{3},l_{9}\right)$
(i.e., interference due to user 1's DSs ). 

Thus, at $t=4,5$, the transmitter sends $2$ useful linear combinations
$l_{8}$ and $l_{9}$ to $R_{1}$, the composite interference symbols
$l_{2}+n_{1}$ and $l_{3}+m_{2}$ to receivers $R_{2}$ and $R_{3}$
respectively, which then use this knowledge to cancel out the interference
from $Y_{2}(1)$ and $Y_{3}(1)$ to obtain their respective desired
linear combinations $m_{1}$ and $n_{2}$. At the same time, $R_{2}$
and $R_{3}$ are also provided with their other desired linear combination
i.e., $m_{2}$ and $n_{1}$, albeit with additional interference $k_{2}$
and $k_{1}$ respectively.

At time $t=6,$ the transmitter transmits $k_{1}$ and $l_{6}+m_{4}$,
the latter in the null space of the channel to $R_{1}$ at $t=6$,
resulting in $R_{1}$ seeing $k_{1}$ and doing layer peeling on it
i.e., canceling out $l_{8}$, to obtain its desired LC $l_{2}$. $R_{2}$
also cancels out the contribution of $l_{8}$ (which it knows from
$t=3$) from $k_{1}$, and thus obtains a linear combination of $l_{2},l_{6}$
and $m_{4}$, in which we group the interference due to user 1 DSs
into the auxiliary symbol $k_{3}=\left(l_{2},l_{6}\right)$, a linear
combination of $l_{2}$ and $l_{6}$, and thus $Y_{2}(6)=\left(k_{3},m_{4}\right)$
. $R_{3}$ sees a linear combination of $k_{1}$ and $l_{6}+m_{4}$,
which we have shown separately in Fig. \ref{fig:-Alternating-CSIT-Example}
to emphasize the fact that $R_{3}$ sees $l_{6}+m_{4}$ with the same
interference $k_{1}$ that it encountered at $t=4$. 

At time $t=7,$ the transmitter similarly sends $k_{2}$ and $l_{5}+n_{3}$,
the latter in the null space of the channel to $R_{1}$ at $t=6$,
resulting in $R_{1}$ seeing $k_{2}$ and canceling out $l_{9}$,
to obtain its desired LC $l_{3}$. $R_{3}$ also cancels out the contribution
of $l_{9}$ (which it knows from $t=3$) from $k_{2}$, and thus obtains
a linear combination of $l_{3},l_{5}$ and $n_{3}$, in which we group
the interference due to user 1 DSs into the auxiliary symbol $k_{4}=\left(l_{3},l_{5}\right)$,
and thus $Y_{3}(7)=\left(k_{3},n_{3}\right)$ . $R_{2}$ sees a linear
combination of $k_{2}$ and $l_{5}+n_{3}$, which we have again shown
separately to emphasize that $R_{3}$ sees $l_{5}+n_{3}$ with the
same interference $k_{2}$ that it encountered at $t=5$. 

Thus, at $t=6,7$, $R_{1}$ obtains two useful linear combinations
$l_{2}$ and $l_{3}$, while both $R_{2}$ and $R_{3}$ acquire one
desired LC each i.e., $m_{4}$ and $n_{3}$, but with added interference
$k_{3}$ and $k_{4}$ respectively. We note that layer peeling is
once again the motivation behind creating the auxiliary interference
symbols $k_{3}$ and $k_{4}$, e.g. in the subsequent NPP phase, $k_{3}$
is useful in its entirety at $R_{3}$ and also manages to deliver
the desired LC $l_{6}$ at $R_{1}$. The other major idea is to repeat
the previously seen interference $k_{1}$ at $R_{3}$ (at $t=5$)
and $k_{2}$ at $R_{2}$ (at $t=6)$, respectively. This sets up the
endgame where knowledge of $k_{1}$ at $R_{3}$ will lead simultaneously
to interference cancellation from $Y_{3}(4)$ which provides $n_{1}$
at $R_{3}$ and the unwrapping of the serial alignment chain $k_{1}\rightarrow l_{6}+m_{4}\rightarrow n_{4}$
which furnishes $n_{4}$ at $R_{3}$. A similar logic shows that knowledge
of $k_{2}$ at $R_{2}$ in the subsequent phase allows $R_{2}$ to
acquire $m_{3}$ and $m_{4}$. This is our motivation to switch to
the NPP CSIT mode for the next $2$ time slots, and use the perfect
knowledge of channels to $R_{2}$ and $R_{3}$ to deliver the auxiliary
interference symbols $k_{2},k_{3}$ to $R_{2}$ and $k_{1},k_{4}$
to $R_{3}$ without any interference.

\emph{\uline{NPP Phase}}

At $t=8$, the transmitter sends $k_{1}$ and $k_{3}$, in the null
space of the channels to $R_{2}$ and $R_{3}$ at $t=8$ respectively
(possible due to the now instantaneous knowledge about the channels
from the transmitter to $R_{2}$ and $R_{3}$). $R_{1}$ sees a linear
combination of $k_{1}$ and $k_{3}$ , from which it cancels out the
contribution of $l_{2}$(known previously at $t=6$) and $l_{8}$
(known at $t=4$) to obtain its required DS $l_{6}$. Since $k_{1}$
is zero-forced at $R_{2}$, $R_{2}$ sees only $k_{3}$, and uses
it to cancel the interference from $Y_{2}(6)$ to obtain its desired
LC $m_{4}$. $R_{3}$ is able to acquire $k_{1}$ without any interference,
because of zero-forcing of $k_{3}$ at $R_{3}$. As discussed in the
previous paragraph, knowledge of $k_{1}$ lets $R_{3}$ cancel interference
from $Y_{3}(4)$ to obtain its desired LC $n_{1}$ and also allows
$R_{3}$ to do interference cancellation from $Y_{3}(6)$ to obtain
$l_{6}+m_{4}$, which in turn allows interference cancellation from
$Y_{3}(2)$ to provide $R_{3}$ with another desired LC $n_{4}$,
a chain of events that is elegantly described by the serial alignment
chain $k_{1}\rightarrow l_{6}+m_{4}\rightarrow n_{4}$. 

In the final time slot $t=9$, the transmitter sends $k_{4}$ and
$k_{2}$, respectively in the null space of the channels to $R_{2}$
and $R_{3}$ at $t=9$. $R_{1}$ uses its previous knowledge of $l_{3}$
and $l_{9}$ to cancel out their contribution from the linear combination
of $k_{2}$ and $k_{4}$ to obtain its final required LC $l_{5}$.
$R_{2}$, which sees only $k_{2}$ free from interference, uses $k_{2}$
to cancel interference from $Y_{2}(5)$ to obtain the desired LC $m_{2}$
and also to unravel the serial alignment chain $k_{2}\rightarrow l_{5}+n_{3}\rightarrow m_{3}$
(from $Y_{3}(9)$,$Y_{3}(7)$ and $Y_{3}(2)$) to obtain its final
desired LC $m_{3}$. $R_{3}$ sees an interference free version of
$k_{4}$, and it cancels its contribution from $Y_{3}(7)$ to obtain
the LC $n_{3}$.

At the end of the $9$ time slots, $R_{1}$ has the LCs $l_{1}$,\ldots{},$l_{9}$,
and can thus decode its DSs $u_{1}$,\ldots{},$u_{9}$. $R_{2}$
possesses the LCs $m_{1}$,\ldots{},$m_{4}$ which enable it to decode
all its desired DSs $v_{1}$,\ldots{},$v_{4}$, and similarly, $R_{3}$
uses the LCs $n_{1}$,\ldots{},$n_{4}$ to decode its DSs $w_{1}$,\ldots{},$w_{4}$.

\section{Conclusion \label{sec:Conclusion}}

In this paper, we obtain an outer bound for the DoF region of the
K-user MISO BC in the most general hybrid CSIT setting, where an arbitrary
number of receivers are in the instantaneous CSIT mode and the rest
of the receivers are in the delayed CSIT mode. We specialize these
results for the 3-user MISO BC, where the transmitter has instantaneous
CSIT about one receiver and delayed CSIT about the other two. We develop
new communication schemes for the 3-user BC for the hybrid CSIT model,
and demonstrate achievability of a sum-DoF that is more than that
obtainable only with delayed CSIT. We also show how an outer bound
corner-point for the hybrid CSIT can be achieved using alternating
CSIT.

\bibliographystyle{IEEEtran}
\bibliography{Bibliography,bibliography}

\end{document}

%% file: dofregion_3.pdf_tex
\begingroup%
  \makeatletter%
  \providecommand\color[2][]{%
    \errmessage{(Inkscape) Color is used for the text in Inkscape, but the package 'color.sty' is not loaded}%
    \renewcommand\color[2][]{}%
  }%
  \providecommand\transparent[1]{%
    \errmessage{(Inkscape) Transparency is used (non-zero) for the text in Inkscape, but the package 'transparent.sty' is not loaded}%
    \renewcommand\transparent[1]{}%
  }%
  \providecommand\rotatebox[2]{#2}%
  \ifx\svgwidth\undefined%
    \setlength{\unitlength}{674.44658203bp}%
    \ifx\svgscale\undefined%
      \relax%
    \else%
      \setlength{\unitlength}{\unitlength * \real{\svgscale}}%
    \fi%
  \else%
    \setlength{\unitlength}{\svgwidth}%
  \fi%
  \global\let\svgwidth\undefined%
  \global\let\svgscale\undefined%
  \makeatother%
  \begin{picture}(1,0.60999723)%
    \put(0,0){\includegraphics[width=\unitlength]{dofregion_3.pdf}}%
    \put(0.43270759,0.28423918){\makebox(0,0)[lt]{\begin{minipage}{0.27939436\unitlength}\raggedright  $\frac{d_1}{3}+\frac{d_3}{2}+d_2\leq 1$\end{minipage}}}%
    \put(0.36238384,0.42678729){\makebox(0,0)[lt]{\begin{minipage}{0.27939436\unitlength}\raggedright  $\frac{d_1}{3}+\frac{d_2}{2}+d_3\leq 1$\end{minipage}}}%
    \put(0.40229732,0.17780324){\makebox(0,0)[lt]{\begin{minipage}{0.27939436\unitlength}\raggedright $\frac{d_1}{2}+d_2\leq 1$\end{minipage}}}%
    \put(0.20463059,0.49140912){\makebox(0,0)[lt]{\begin{minipage}{0.27939436\unitlength}\raggedright $\frac{d_1}{2}+d_3\leq 1$\end{minipage}}}%
    \put(0.324371,0.59784503){\makebox(0,0)[lt]{\begin{minipage}{0.04561545\unitlength}\raggedright $d_1$\end{minipage}}}%
    \put(0.10070792,0.31798954){\makebox(0,0)[lt]{\begin{minipage}{0.04561545\unitlength}\raggedright $d_3$\end{minipage}}}%
    \put(0.30634289,0.05625766){\makebox(0,0)[lt]{\begin{minipage}{0.04561545\unitlength}\raggedright $d_2$\end{minipage}}}%
    \put(0.16707546,0.56030387){\makebox(0,0)[lt]{\begin{minipage}{0.04561545\unitlength}\raggedright 1\end{minipage}}}%
    \put(0.39849603,0.58359719){\makebox(0,0)[lt]{\begin{minipage}{0.04561545\unitlength}\raggedright 0\end{minipage}}}%
    \put(0.15046925,0.54132535){\makebox(0,0)[lt]{\begin{minipage}{0.04561545\unitlength}\raggedright 1\end{minipage}}}%
    \put(0.17231967,0.14454896){\makebox(0,0)[lt]{\begin{minipage}{0.04561545\unitlength}\raggedright 0\end{minipage}}}%
    \put(0.50969052,0.04145658){\makebox(0,0)[lt]{\begin{minipage}{0.04561545\unitlength}\raggedright 1\end{minipage}}}%
    \put(0.18942544,0.1236419){\makebox(0,0)[lt]{\begin{minipage}{0.04561545\unitlength}\raggedright 0\end{minipage}}}%
    \put(0.15584507,0.33838789){\makebox(0,0)[lt]{\begin{minipage}{0.04561545\unitlength}\raggedright $\frac{1}{2}$\end{minipage}}}%
    \put(0.343999,0.07900426){\makebox(0,0)[lt]{\begin{minipage}{0.04561545\unitlength}\raggedright  $\frac{1}{2}$\end{minipage}}}%
    \put(0.61485758,0.435){\makebox(0,0)[lt]{\begin{minipage}{0.13779643\unitlength}\raggedright $(0,\frac{2}{3},\frac{2}{3})$\end{minipage}}}%
    \put(0.23,0.26){\makebox(0,0)[lt]{\begin{minipage}{0.13779643\unitlength}\raggedright $(1,\frac{4}{9},\frac{4}{9})$\end{minipage}}}%
    \put(0.25,0.21){\makebox(0,0)[lt]{\begin{minipage}{0.13779643\unitlength}\raggedright $(1,\frac{1}{2},\frac{1}{3})$\end{minipage}}}%
    \put(0.2,0.30790395){\makebox(0,0)[lt]{\begin{minipage}{0.13779643\unitlength}\raggedright $(1,\frac{1}{3},\frac{1}{2})$\end{minipage}}}%
  \end{picture}%
\endgroup%

%% file: K-user_MISO_BC_with_Hybrid_CSIT__v4_.bbl
\begin{thebibliography}{10}
\providecommand{\url}[1]{#1}
\csname url@samestyle\endcsname
\providecommand{\newblock}{\relax}
\providecommand{\bibinfo}[2]{#2}
\providecommand{\BIBentrySTDinterwordspacing}{\spaceskip=0pt\relax}
\providecommand{\BIBentryALTinterwordstretchfactor}{4}
\providecommand{\BIBentryALTinterwordspacing}{\spaceskip=\fontdimen2\font plus
\BIBentryALTinterwordstretchfactor\fontdimen3\font minus
  \fontdimen4\font\relax}
\providecommand{\BIBforeignlanguage}[2]{{%
\expandafter\ifx\csname l@#1\endcsname\relax
\typeout{** WARNING: IEEEtran.bst: No hyphenation pattern has been}%
\typeout{** loaded for the language `#1'. Using the pattern for}%
\typeout{** the default language instead.}%
\else
\language=\csname l@#1\endcsname
\fi
#2}}
\providecommand{\BIBdecl}{\relax}
\BIBdecl

\bibitem{Weingarten2006}
H.~Weingarten, Y.~Steinberg, and S.~Shamai, ``The capacity region of the
  gaussian multiple-input multiple-output broadcast channel,''
  \emph{Information Theory, IEEE Transactions on}, vol.~52, no.~9, pp. 3936
  --3964, sept. 2006.

\bibitem{Huang2012}
C.~Huang, S.~A. Jafar, S.~Shamai, and S.~Vishwanath, ``On degrees of freedom
  region of {M}{I}{M}{O} networks without channel state information at
  transmitters,'' \emph{IEEE Trans. Inform. Th.}, vol.~58, no.~2, pp. 849--857,
  Feb. 2012.

\bibitem{Vaze2012}
C.~S. Vaze and M.~K. Varanasi, ``The degrees of freedom regions of {M}{I}{M}{O}
  broadcast, interference, and cognitive radio channels with no {C}{S}{I}{T},''
  \emph{IEEE Trans. Inform. Th.}, vol.~58, no.~8, pp. 5354--5374, Aug. 2012.

\bibitem{Maddah-Ali2012}
M.~A. Maddah-Ali and D.~Tse, ``Completely stale transmitter channel state
  information is still very useful,'' \emph{IEEE Trans. Inform. Th.}, vol.~58,
  no.~7, pp. 4418--4431, Jul. 2012.

\bibitem{Vaze2011}
C.~Vaze and M.~Varanasi, ``The degrees of freedom region of the two-user
  {M}{I}{M}{O} broadcast channel with delayed {C}{S}{I}{T},'' in
  \emph{Information Theory Proceedings (ISIT), 2011 IEEE International
  Symposium on}, 31 2011-aug. 5 2011, pp. 199 --203.

\bibitem{Abdoli2011}
M.~J. Abdoli, A.~Ghasemi, and A.~K. Khandani, ``On the degrees of freedom of
  three-user {M}{I}{M}{O} broadcast channel with delayed {C}{S}{I}{T},'' in
  \emph{ISIT}, Aug. 2011, pp. 209--213.

\bibitem{Yang2012}
\BIBentryALTinterwordspacing
S.~Yang, M.~Kobayashi, D.~Gesbert, and X.~Yi, ``Degrees of freedom of time
  correlated {M}{I}{S}{O} broadcast channel with delayed {C}{S}{I}{T},''
  \emph{submitted, IEEE Trans. Inform. Th.}, Mar. 2012. [Online]. Available:
  \url{http://arxiv.org/abs/1203.2550.}
\BIBentrySTDinterwordspacing

\bibitem{Gou2012}
T.~Gou and S.~Jafar, ``Optimal use of current and outdated channel state
  information: Degrees of freedom of the {M}{I}{S}{O} {B}{C} with {M}ixed
  {C}{S}{I}{T},'' \emph{Communications Letters, IEEE}, vol.~16, no.~7, pp. 1084
  --1087, Jul. 2012.

\bibitem{Chen2012}
\BIBentryALTinterwordspacing
J.~Chen and P.~Elia, ``Degrees-of-freedom region of the {M}{I}{S}{O} broadcast
  channel with general mixed-{C}{S}{I}{T},'' May 2012. [Online]. Available:
  \url{http://arxiv.org/abs/1205.3474.}
\BIBentrySTDinterwordspacing

\bibitem{Tandon2012}
\BIBentryALTinterwordspacing
R.~Tandon, S.~A. Jafar, S.~Shamai, and H.~V. Poor, ``On the synergistic
  benefits of alternating {C}{S}{I}{T} for the {M}{I}{S}{O} {B}{C},'' Aug.
  2012. [Online]. Available: \url{http://arxiv.org/abs/1208.5071.}
\BIBentrySTDinterwordspacing

\bibitem{Tandon2012a}
R.~Tandon, M.~A. Maddah-Ali, A.~Tulino, H.~V. Poor, and S.~Shamai, ``On fading
  broadcast channels with partial channel state information at the
  transmitter,'' in \emph{Wireless Communication Systems (ISWCS), 2012
  International Symposium on}, aug. 2012, pp. 1004 --1008.

\bibitem{Mohanty2012}
\BIBentryALTinterwordspacing
K.~Mohanty, C.~S. Vaze, and M.~K. Varanasi, ``The degrees of freedom region for
  the {M}{I}{M}{O} interference channel with hybrid {C}{S}{I}{T},'' Sep. 2012.
  [Online]. Available: \url{http://arxiv.org/abs/1209.0047.}
\BIBentrySTDinterwordspacing

\bibitem{Rassouli2013}
\BIBentryALTinterwordspacing
B.~Rassouli, C.~Hao, and B.~Clerckx, ``{D}o{F} analysis of the {K}-user
  {M}{I}{S}{O} broadcast channel with alternating {C}{S}{I}{T},'' Nov. 2013.
  [Online]. Available: \url{http://arxiv.org/abs/1311.6647.}
\BIBentrySTDinterwordspacing

\bibitem{Gamal1978}
A.~Gamal, ``The feedback capacity of degraded broadcast channels (corresp.),''
  \emph{Information Theory, IEEE Transactions on}, vol.~24, no.~3, pp. 379 --
  381, may 1978.

\end{thebibliography}
